\documentclass[aps,pre,12pt,notitlepage,tightenlines]{revtex4-1}

\usepackage{amsmath}
\usepackage{amsfonts}
\usepackage{bm}
\usepackage{euscript}
\usepackage{graphicx}
\usepackage{wasysym}
\usepackage{subfigure}
\usepackage{ifthen}

\newcommand{\APS}{false}
\newcommand{\ifaps}[2]{\ifthenelse{\equal{\APS}{true}}{#1}{#2}}

\ifaps{}{\pdfoutput=1}

\usepackage[today,nofancy]{svninfo}
\svnInfo $Id: movingwall.tex 163 2011-08-25 16:48:43Z jeanluc $

\ifaps{\graphicspath{{figs_eps/}}}{\graphicspath{{figs/}{figs_lo/}}}

\newcommand{\mathnotation}[2]{\newcommand{#1}{\ensuremath{#2}}}

\newcommand{\Order}[1]{\mathrm{O}\!\l(#1\r)}

\newcommand{\coloronline}{\ifaps{(Color online)\ }{}}

\renewcommand{\l}{\left}               % \left
\renewcommand{\r}{\right}              % \right
\mathnotation{\pd}{\partial}           % Partial derivative
\mathnotation{\ee}{{\mathrm e}}        % e
\mathnotation{\dint}{\,{\mathrm{d}}}   % Differential, in an integral

\renewcommand{\time}{t}                % Time
\mathnotation{\xc}{\theta}             % Angle around container
\mathnotation{\yc}{y}                  % Spatial coordinate y
\mathnotation{\xcp}{\bar\xc}           % Map iterate x
\mathnotation{\ycp}{\bar\yc}           % Map iterate y
\mathnotation{\xcsep}{\xc_{\mathrm{s}}}% Stable separatrix
\mathnotation{\ycsep}{\yc_{\text{s}}}  % Separatrix equation
\mathnotation{\uc}{u}                  % Stirring velocity field x component
\mathnotation{\vc}{v}                  % Stirring velocity field y component
\mathnotation{\Uc}{\Omega}             % Wall velocity
\mathnotation{\A}{A}                   % Near-wall velocity
\mathnotation{\Xc}{\Theta}             % Linearised coordinate
\mathnotation{\Yc}{Y}                  % Linearised coordinate
\mathnotation{\Xcp}{\overline{\Xc}}    % Linearised map iterate x
\mathnotation{\Ycp}{\overline{\Yc}}    % Linearised map iterate y
\mathnotation{\T}{T}                   % Period
\mathnotation{\decfp}{\mu}             % Rate of decay to the fixed point
\mathnotation{\decvar}{\alpha}         % Variance decay rate
\mathnotation{\lyap}{\lambda}          % Typical stretching rate
\mathnotation{\gap}{d}                 % Width of gap
\mathnotation{\dotgap}{\skew{8}\dot{d}}% Rate of change of \gap
\mathnotation{\lB}{\ell_{\mathrm{B}}}  % Batchelor scale
\mathnotation{\tB}{\time_{\mathrm{B}}} % Time to reach Batchelor scale
\mathnotation{\Aw}{\mathcal{A}_{\mathrm{w}}}% Area of visitble white material
\mathnotation{\Sgen}{S}                % Map generating function
\mathnotation{\Fgen}{F}                % Map generating function
\mathnotation{\aaa}{a}                 % Near-wall flow velocity
\mathnotation{\Ham}{H}                 % Hamiltonian

\begin{document}

\title{Moving walls accelerate mixing}

\author{Jean-Luc Thiffeault}
\affiliation{Department of Mathematics, University of Wisconsin --
  Madison, WI 53706, USA}
\author{Emmanuelle Gouillart}
\affiliation{Surface du Verre et Interfaces, UMR 125 CNRS/Saint-Gobain, 93303
Aubervilliers, France}
\author{Olivier Dauchot}
\affiliation{Service de Physique de l'Etat Condens\'e, DSM, CEA Saclay,
URA2464, 91191 Gif-sur-Yvette Cedex, France}

\date{\today}

%\pacs{47.27.Qb, % Turbulent diffusion
%92.10.Lq, %Turbulence and diffusion
%92.60.Ek, %Convection, turbulence and diffusion
%94.10.Lf %convection, diffusion, mixing turbulence and fallout
%}
%\keywords{advection, convection, diffusion, mixing, turbulent diffusion}

\begin{abstract}
  Mixing in viscous fluids is challenging, but chaotic advection in
  principle allows efficient mixing.  In the best possible scenario,
  the decay rate of the concentration profile of a passive scalar
  should be exponential in time.  In practice, several authors have
  found that the no-slip boundary condition at the walls of a vessel
  can slow down mixing considerably, turning an exponential decay into
  a power law.  This slowdown affects the whole mixing region, and not
  just the vicinity of the wall.  The reason is that when the chaotic
  mixing region extends to the wall, a separatrix connects to it.  The
  approach to the wall along that separatrix is polynomial in time and
  dominates the long-time decay.  However, if the walls are moved or
  rotated, closed orbits appear, separated from the central mixing
  region by a hyperbolic fixed point with a homoclinic orbit.  The
  long-time approach to the fixed point is exponential, so an overall
  exponential decay is recovered, albeit with a thin unmixed region
  near the wall.
\end{abstract}

\maketitle

\section{Introduction}
\label{sec:intro}

In many engineering applications, a viscous fluid must be blended with
a substance, referred to as a passive scalar.  This is generally
called mixing, and it is well known that stirring greatly enhances
this process.  In fact, even if turbulence is unavailable (because of
low Reynolds number or delicate substances), it is possible to mix
rapidly by the process of chaotic advection~\cite{Aref1984,Ottino}.
This involves the chaotic stretching of fluid particles by the flow,
and the subsequent increase in concentration gradients of the
substance to be mixed.  The increased gradients facilitate the action
of molecular diffusion, and homogeneity ensues.  The net process
(chaotic advection together with diffusion) is known as chaotic
mixing.

\begin{figure}
\subfigure[]{
  \ifaps{
    \includegraphics[width=.4\textwidth]{fig1a}
  }{
    \includegraphics[width=.4\textwidth]{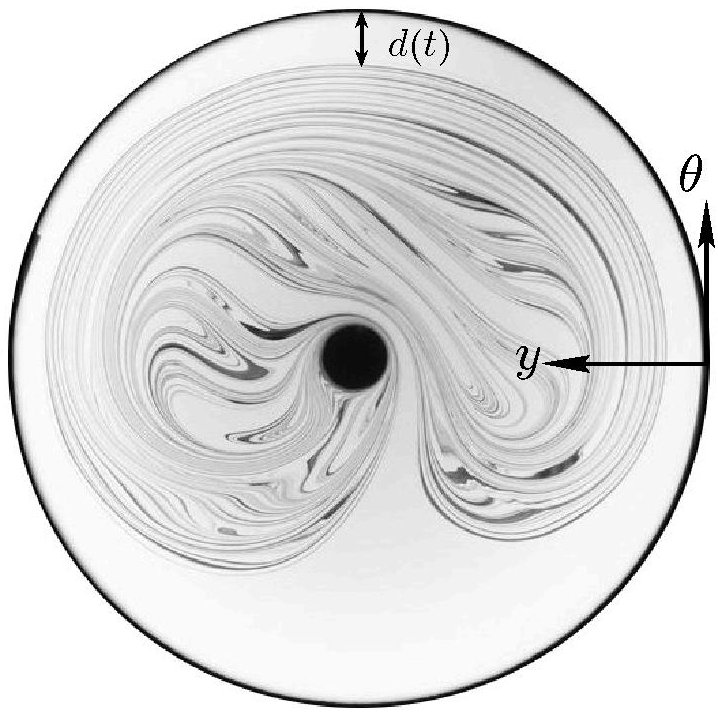}
  }
  \label{fig:fig8exp}
}\hspace{1em}%
\subfigure[]{
  \ifaps{
    \includegraphics[width=.43\textwidth]{fig1b}
  }{
    \includegraphics[width=.43\textwidth]{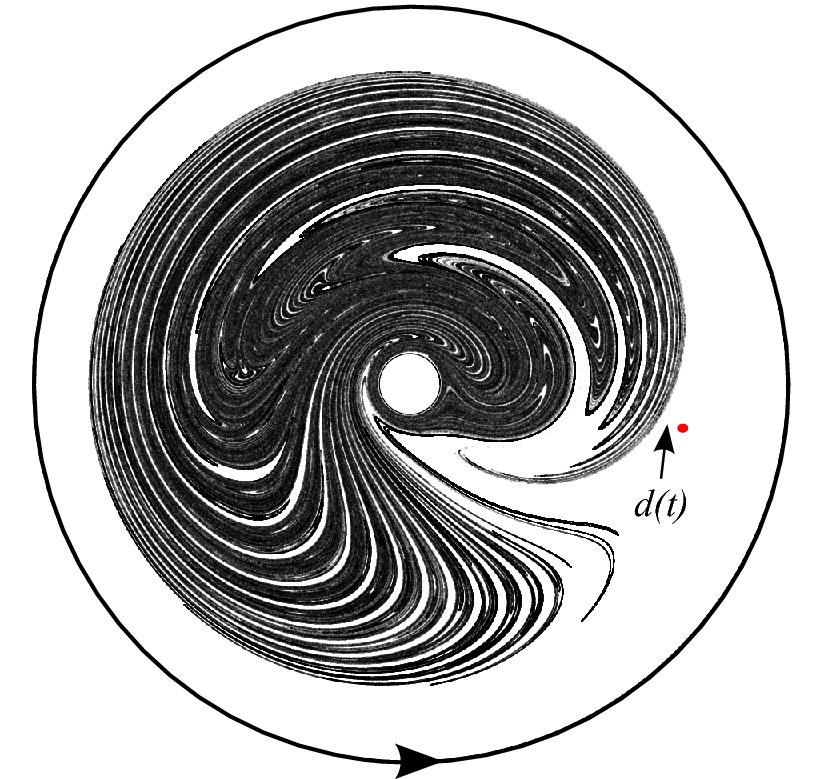}
  }
  \label{fig:fig8rot}
}
\caption{(a) Experiment with the `figure-eight' stirring protocol,
  showing an advected blob of dye (India ink) in sugar syrup (details
  of the experimental apparatus are given
  in~\cite{Gouillart2007,Gouillart2008}).  The coordinate system used
  here is also indicated, as is the distance~$\gap(\time)$ between the
  dark mixing pattern and the distinguished parabolic point on the
  wall.  See Fig.~\ref{fig:eightpoinc} for the rod's path. (b)
  Numerical simulation for the same stirring protocol as in (a), but
  with the wall rotating at a rate~$\Omega=0.4\ \mathrm{rad/s}$.
  Here,~$\gap(\time)$ is the distance from the mixing pattern to the
  hyperbolic point indicated by a dot.}
\label{fig:fig8both}
\end{figure}

The quantity that is often tracked in mixing problems is the variance
of the concentration of the scalar~\cite{Danckwerts1952,
  Rothstein1999, Jullien2000, Fereday2002, Sukhatme2002, Voth2003,
  Villermaux2003, Thiffeault2003d}. The variance is the spatial
integral of the squared deviation from the mean concentration, and
measures therefore the intensity of concentration fluctuations. The
reasoning is that the variance tends to zero as the concentration is
homogenized.  Simple arguments~\cite{Zeldovich1984, Shraiman1994,
  Antonsen1996, Balkovsky1999, Fereday2002, Thiffeault2003d,
  ThiffeaultAosta2004, Haynes2005, Meunier2010} suggest that the
concentration variance should decay exponentially with time.  In an
idealized scenario the fluid particles are stretched exponentially and
folded by the flow, yielding a characteristic filamentary structure as
in Fig.~\ref{fig:fig8both}.  The filaments then achieves an
equilibrium width where diffusion balances
stretching~\cite{Batchelor1959}. Subsequently, the concentration field
for that particle decays exponentially at the rate of stretching.  An
average over rates of stretching then gives the overall decay rate of
the variance.  In many cases the decay rate of the variance is
determined in a less local manner~\cite{Pierrehumbert1994,
  Fereday2002, Wonhas2002, Pikovsky2003, Fereday2004, Thiffeault2003d,
  Haynes2005}, but the decay is still exponential.

This basic exponential-decay picture is appealing, but it is
complicated by the presence of walls.  In that case, several
authors~\cite{Jones1994, Chertkov2003, Lebedev2004, Schekochihin2004,
  Salman2007, Popovych2007, Chernykh2008, Mackay_CCT2007,
  Boffetta2009, Zaggout2011} have suggested that the no-slip boundary
condition and the presence of separatrices on the walls slow down
mixing: the decay rate is algebraic rather than exponential.  Recent
experiments~\cite{Gouillart2007,Gouillart2008,Gouillart2009} have
confirmed this, and also showed that for a significant period of time
the rate of decay of variance is dramatically reduced, even away from
the walls, due to the entrainment of unmixed material into the central
mixing region.

In this paper, which is a more detailed and complete exposition of an
earlier letter~\cite{Gouillart2010}, we show that if the situation is
such that mixing is slowed down by no-slip walls, then this can be
cured by moving the wall in such a way as to destroy the separatrices.
This creates closed orbits near the wall, effectively insulating the
central mixing region from the wall (see
Fig.~\ref{fig:eightpoinc}). The rate of decay of the concentration
variance becomes exponential rather than algebraic.  The price to pay
is that the thin region of closed orbits remains poorly mixed.  Of
course, in some applications moving or spinning the outer wall of a
container is not practical, so we also discuss other mechanisms for
creating closed orbits near the wall.

The outline of the paper is as follows.  In Section~\ref{sec:fixed} we
analyze the case of a device consisting of a stirring rod and a fixed
outer wall.  We describe the separatrices that appear at the boundary,
and how the approach along the stable manifold of one of these
separatrices is algebraic in time.  In Section~\ref{sec:movingwall} we
treat the same system, but with the wall rotating at a constant
velocity.  We find the wall separatrices are destroyed; instead, a
hyperbolic fixed point appears away from the wall, with a homoclinic
orbit separating the wall region from the central mixing region.  The
approach along the stable manifold of that fixed point is exponential
in time.

Section~\ref{sec:var} relates these results to the decay of
concentration variance.  We show that if the wall is rotating fast
enough, the decay rate is not limited by the no-slip boundary
condition, and mixing proceeds as rapidly as it would in the absence
of a no-slip wall.  We present numerical evidence for this in
Section~\ref{sec:simulation}, based on simulations of the evolution of
the concentration field for a simple, but realistic, flow.  We then
offer some concluding remarks, and discuss other ways of creating
closed orbits beyond physically rotating the wall.

\section{Fixed Wall}
\label{sec:fixed}

Consider the experiment for the time-periodic `figure-eight'
rod-stirring protocol shown in Fig.~\ref{fig:fig8exp}.  The
numerically-computed Poincar\'e section
\begin{figure}
\subfigure[]{
  \ifaps{
    \includegraphics[width=.4\textwidth]{fig2a}
  }{
    \includegraphics[width=.4\textwidth]{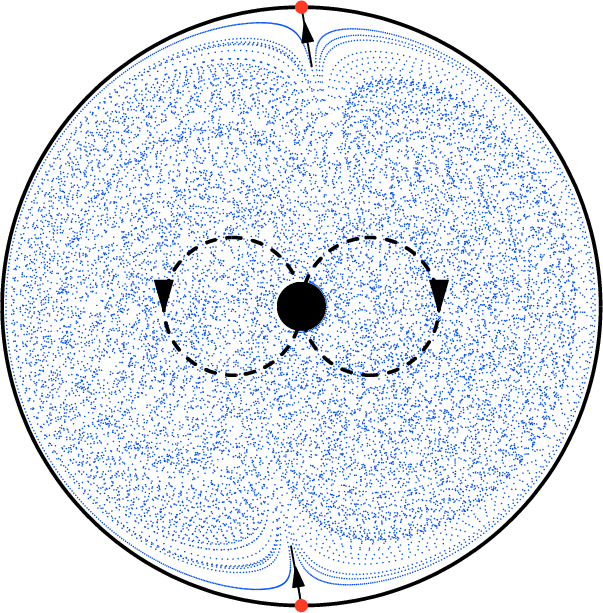}
  }
  \label{fig:eightpoinc_wout=0}
}\hspace{2em}
\subfigure[]{
  \ifaps{
    \includegraphics[width=.4\textwidth]{fig2b}
  }{
    \includegraphics[width=.4\textwidth]{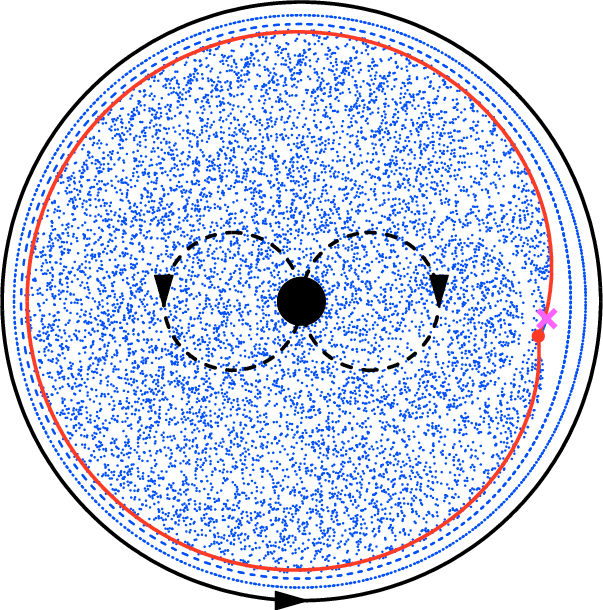}
  }
  \label{fig:eightpoinc_wout=0p2}
}
\caption{\protect\coloronline (a) Numerical Poincar\'e section
  (stroboscopic map) of a single trajectory for the time-periodic
  figure-eight stirring protocol (dashed line).  The circular domain
  has unit radius, the rod has radius~$0.08$, and the rod's two
  circular paths have radius~$0.23$.  The chaotic region covers the
  entire domain, all the way to the wall, where two near-wall
  separatrices are indicated in the upper and lower portions of the
  domain. (b) Same parameters, but with the outer wall rotating at a
  rate of~$\Uc=0.2$ radians per period.  Closed orbits appear near the
  wall, and the wall separatrices are destroyed.  There is however a
  separatrix associated with a hyperbolic fixed point of the map
  (solid line), which isolates the chaotic region from the wall.  The
  solid dot is the approximate position of the fixed point according
  to the model in Section~\ref{sec:movingwall}, and the cross is the
  numerically-measured fixed point.  The discrepancy can easily be
  resolved by expanding to next order in~$\Uc$.}
\label{fig:eightpoinc}
\end{figure}%
for the same 2-D Stokes flow (Fig.~\ref{fig:eightpoinc_wout=0}) shows
that this protocol promotes chaotic advection in the whole domain, as
could be expected from the stretching and folding of dye filaments
visible in Fig.~\ref{fig:fig8exp}.  It also shows clearly that
particles move very slowly near the wall, as evidenced by the
closely-packed successive iterates, in accordance with the no-slip
boundary condition.  This suggests writing the flow near the wall as a
simple map,
\begin{equation}
  \xcp = \xc + \A(\xc)\yc + \Order{\yc^2},\qquad
  \ycp = \yc - \tfrac12\A'(\xc)\yc^2 + \Order{\yc^3}.
  \label{eq:nearwall}%
\end{equation}
where~$(\xc,\yc)$ is the position of a fluid particle at the beginning
of a time interval,~$(\xcp,\ycp)$ its new position at the end.
Here~$\xc$ is an angle measured counterclockwise along the wall,
and~$0\le\yc\ll1$ measures the distance from the wall
(Fig.~\ref{fig:fig8exp}).  The first equation in~\eqref{eq:nearwall}
uses continuity and the no-slip boundary condition at the wall; then
the second equation follows from incompressibility and the
no-throughflow condition at the wall.  The container is assumed to
have unit radius.  We can write~$(\xc,\yc)=(\xc(\time),\yc(\time))$
and~$(\xcp,\ycp)=(\xc(\time+\T),\yc(\time+\T))$, where~$\time$ is
time, $\T$ is the period, and~$\time/\T$ is the number of full periods
of the stirring protocol that the system has undergone.  If we
truncate the map by neglecting the higher-order terms, then it
preserves area only to first order in~$\yc$.  We shall see in
Section~\ref{sec:movingwall} how to correct the map to ensure exact
area preservation.

\begin{figure}
  \ifaps{
    \includegraphics[width=.65\textwidth]{fig3}
  }{
    \includegraphics[width=.65\textwidth]{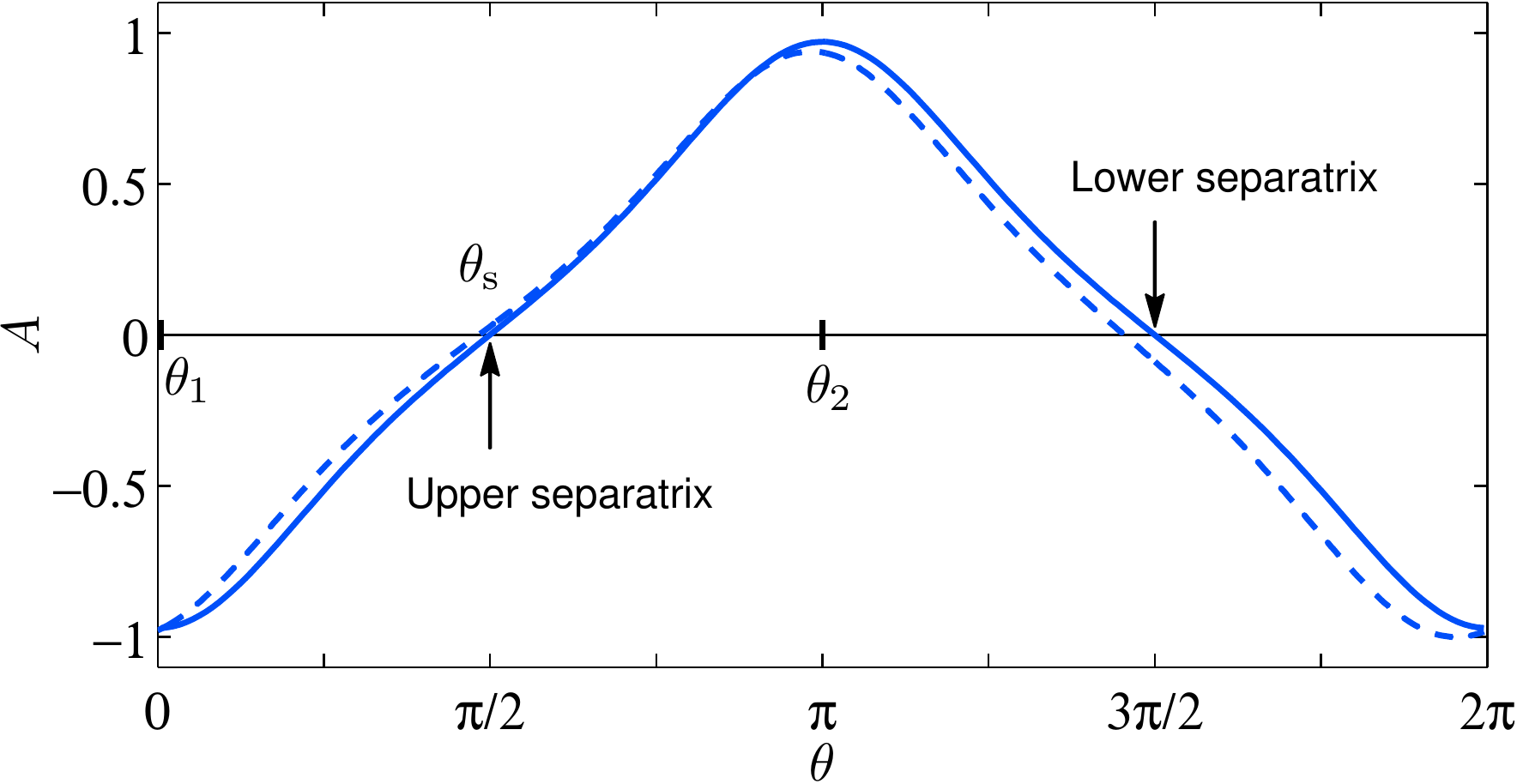}
  }
  \caption{Solid line: The periodic function~$\A(\xc)$ in
    Eq.~\eqref{eq:nearwall} for the figure-eight stirring motion of
    Fig.~\ref{fig:eightpoinc_wout=0}.  The two zeros correspond to the
    separatrices connected to the wall at the points highlighted in
    Fig.~\ref{fig:eightpoinc_wout=0}.  The upper (stable) separatrix
    is at~$\xc=\xcsep=\pi/2$, and the lower (unstable) separatrix is
    at~$\xc=3\pi/2$.  The dashed line is for the same stirring motion,
    but with the wall rotating at angular velocity~$\Uc=0.2$ per
    stirring period.}
  \label{fig:A_eight}
\end{figure}

Since it is continuous, the tangential velocity near the wall can only
reverse sign at points~$\xc$ with~$\A(\xc)=0$.  These points
correspond to the upper and lower wall separatrices (also called
separation and reattachment points) visible in
Fig.~\ref{fig:eightpoinc_wout=0}.  (Note that these only act as
separatrices near the wall, where the flow is almost steady; further
away from the wall the manifolds of the separation points take a more
convoluted form that permits chaotic transport in the bulk of the
vessel.)  There are only two such separatrices, so we conclude
that~$\A(\xc)$ must be as in Fig.~\ref{fig:A_eight}, with the two
zeros
corresponding to the separatrices in Fig.~\ref{fig:eightpoinc_wout=0}.
(The functions in Fig.~\ref{fig:A_eight} were obtained numerically
from 2D Stokes flow simulations~\cite{MattFinn2003}.)

Every point~$(\xc,\yc)=(\xc_0,0)$ on the wall is a fixed point of the
map~\eqref{eq:nearwall}.  Letting~$(\xc,\yc)=(\xc_0+\Xc,0+\Yc)$,
where~$(\Xc,\Yc)$ are small expansion variables, we find the dynamics
of particle trajectories near that point,
\begin{subequations}
\begin{align}
  \Xcp &= \Xc + \A(\xc_0)\Yc + \A'(\xc_0)\Xc\Yc + \Order{\Xc^2\Yc\,,\,\Yc^2},\\
  \Ycp &= \Yc - \tfrac12\A'(\xc_0)\Yc^2 + \Order{\Xc\Yc^2\,,\,\Yc^3}.
\end{align}
\label{eq:nearwalllin}%
\end{subequations}%
The eigenvalues associated with the linearization
of~\eqref{eq:nearwalllin} are both unity, so~$(\xc_0,0)$ is a
distinguished parabolic fixed point~\cite{ReichlTransition}.  Near
most of these points the dynamics are uninteresting: particles just
stream along the wall following~$\Xcp = \Xc + \A(\xc_0)\Yc$, and
approach or recede from the wall depending on the sign
of~$\A'(\xc_0)$.  Eventually we must give up
using~\eqref{eq:nearwalllin}, since the trajectories always leave the
neighborhood of~$(\xc_0,0)$.  However, for the two values of~$\xc$ for
which~$\A(\xc)$ vanishes (see Fig.~\ref{fig:A_eight}), we get
separatrices near the wall.

We focus on the separatrix at~$\xc=\xcsep=\pi/2$,
where~$\A(\xcsep)=0$, and Eqs.~\eqref{eq:nearwalllin} become
\begin{subequations}
\begin{align}
  \Xcp &= \Xc + \A'(\xcsep)\Xc\Yc + \Order{\Xc^2\Yc\,,\,\Yc^2},\\
  \Ycp &= \Yc - \tfrac12\A'(\xcsep)\Yc^2 + \Order{\Xc\Yc^2\,,\,\Yc^3},
\end{align}
\label{eq:nearshsp}%
\end{subequations}%
with~$(\xc,\yc)=(\xcsep+\Xc,0+\Yc)$ and~$(\Xc,\Yc)$ small expansion
variables.  Now the set~\hbox{$\{\Xc=0,\Yc>0\}$} is invariant for
small~$\Yc$ and corresponds to the separatrix, which is the stable
manifold of the fixed point~$(\xcsep,0)$.  The evolution along the stable
manifold is obtained by iterating
\begin{equation}
  \Ycp = \Yc - \tfrac12\A'(\xcsep)\Yc^2,
  \label{eq:logistic}
\end{equation}
for small~$\Yc$.  This is a logistic map, and as we approach the fixed
point at~$\Yc=0$ we expect~$\Yc$ to change very little at each period.
This suggests writing~\eqref{eq:logistic} as a differential equation
for~$\Yc(\time)$,
\begin{equation}
  \frac{\Ycp - \Yc}{\T} \simeq \dot\Yc = -
  \tfrac1{2\T}\A'(\xcsep)\Yc^2,
  \qquad
  {(\Ycp - \Yc)}/{\T} \ll 1.
\end{equation}
The overdot denotes a time derivative.  The solution is
\begin{equation}
  \Yc(\time) = \frac{\Yc(0)}{1 + \tfrac1{2}\A'(\xcsep)\Yc(0)(\time/\T)}\,.
\end{equation}
For this to represent the stable manifold, we require~$\A'(\xcsep)>0$,
as is clear from Fig.~\ref{fig:A_eight}.  (The other separatrix
exhibits finite-time escape to infinity, which takes particles away
from the wall and into the bulk.)  For long times, the rate of
approach is
\begin{equation}
  \Yc(\time) \sim \frac{2}{\A'(\xcsep)}\,(\time/\T)^{-1}\,,\qquad
  \time/\T\gg(\tfrac12\A'(\xcsep)\Yc(0))^{-1}\,.
  \label{eq:asympara}
\end{equation}
The asymptotic form~\eqref{eq:asympara} for~$\Yc(\time)$ is algebraic
and independent of~$\Yc(0)$.  The consequence of this independence
is visible in the upper part of Fig.~\ref{fig:fig8exp}: material lines
`bunch-up' against each other faster than they approach the wall,
thereby forgetting their initial position.

The slow algebraic approach to the wall exhibited in~\eqref{eq:asympara}
was shown in~\cite{Gouillart2007,Gouillart2008,Gouillart2009} to cause a
drastic slowdown of the overall mixing rate in the vessel.  Indeed, even
though the fundamental action of the stirring protocol is chaotic, as
evident in Fig.~\ref{fig:eightpoinc_wout=0}, a pool of unmixed fluid
remains near the wall for very long times. More importantly, this pool
contaminates the entire mixing pattern, all the way to its core, as
unmixed fluid leaks slowly along the unstable manifold of the separation
point at $\xc=3\pi/2$, resulting in the characteristic cusped shape of the
dye pattern in Fig.~\ref{fig:fig8exp}.  This is because there is no
barrier between the wall and the central mixing region.  In the next
section, we will see how we can create such a barrier by moving the
outer wall.

\section{Moving Wall}
\label{sec:movingwall}

Now we consider the case of a slowly-rotating wall, with mixing
pattern as in Fig.~\ref{fig:fig8rot}.  The near-wall map corresponding
to~\eqref{eq:nearwall} is
\begin{equation}
  \xcp = \xc + \Uc + \A(\xc)\yc,\qquad
  \ycp = \yc - \tfrac12\A'(\xc)\yc^2,
  \label{eq:movingwall}
\end{equation}
where~$0 < \Uc \ll 1$ is the angular displacement of the wall per
period, and we have neglected higher-order terms in~$\yc$.  We assume
that~$\Uc$ is small since in that case~$\A(\xc)$ will change little
from the case with~$\Uc=0$, as can be seen by the dashed line in
Fig.~\ref{fig:A_eight}.

\begin{figure}
\subfigure[]{
  \ifaps{
    \includegraphics[width=.4\textwidth]{fig4a}
  }{
    \includegraphics[width=.4\textwidth]{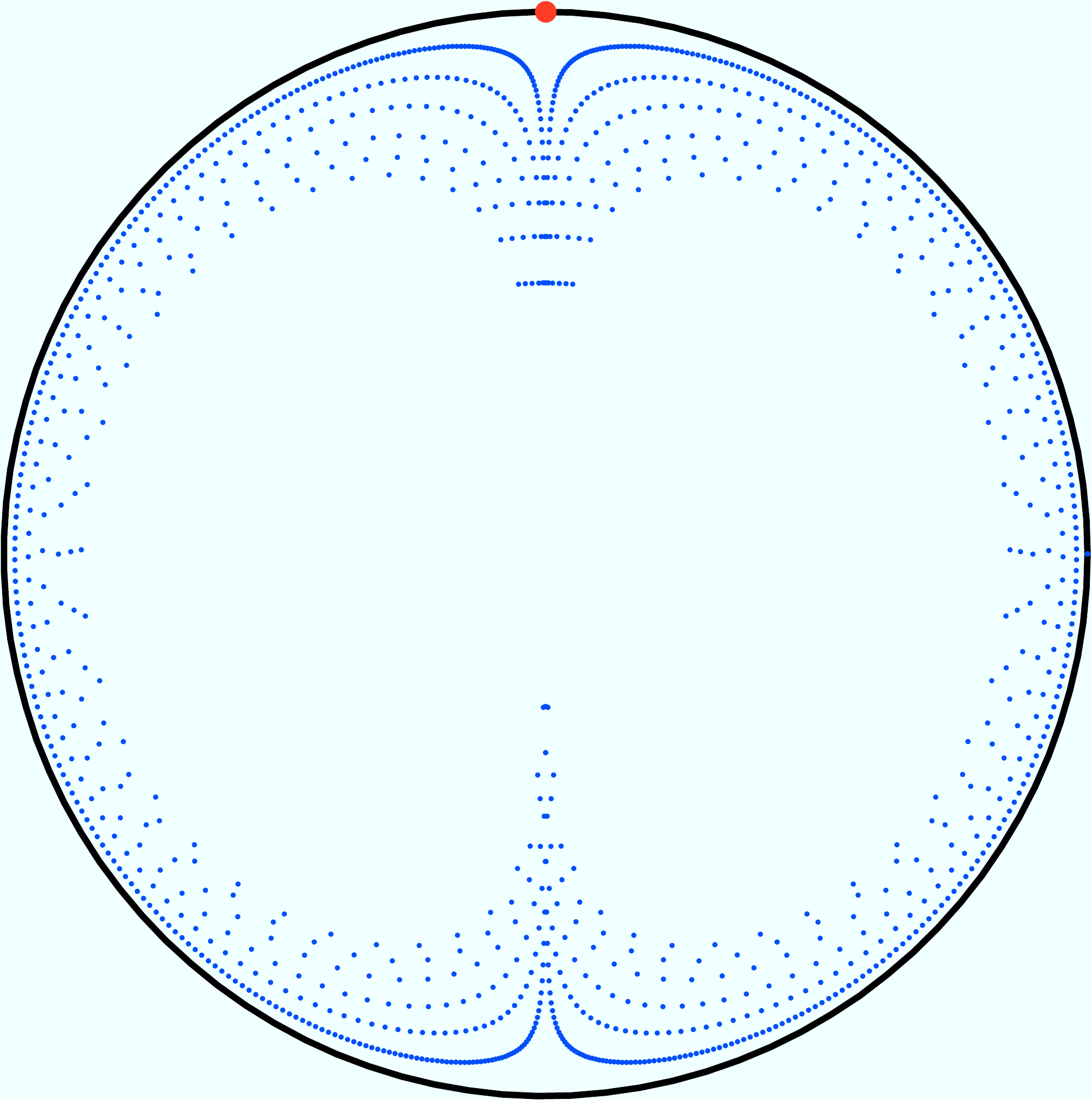}
  }
  \label{fig:lines_U0}
}\hspace{2em}
\subfigure[]{
  \ifaps{
    \includegraphics[width=.4\textwidth]{fig4b}
  }{
    \includegraphics[width=.4\textwidth]{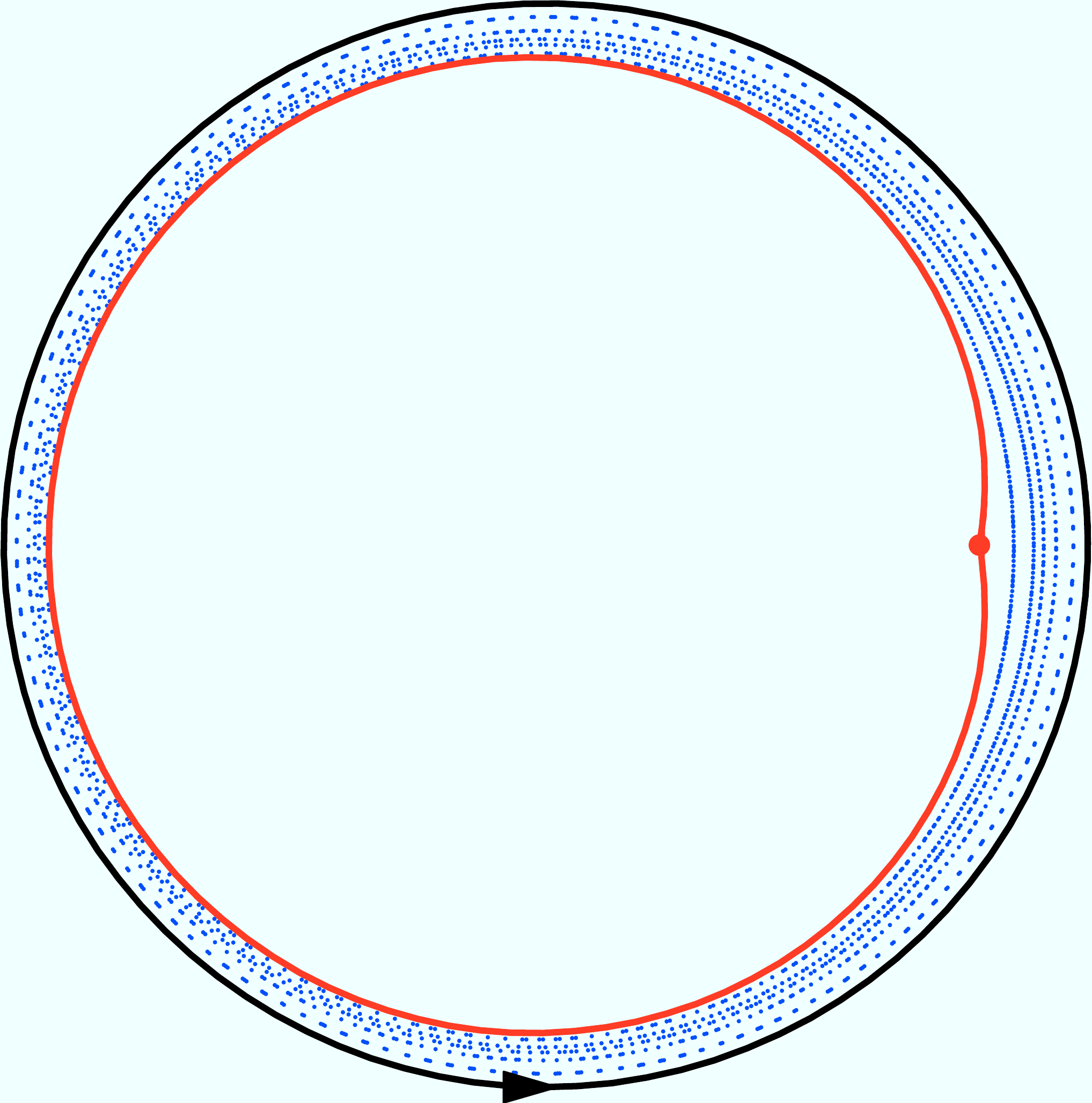}
  }
  \label{fig:lines_U0p2}
}
\caption{\protect\coloronline Iterates the area-preserving near-wall
  map map~\eqref{eq:exactmap}, which should be compared to the full
  numerical simulations of Fig.~\ref{fig:eightpoinc}. (a) $\Uc=0$,
  with the upper stable distinguished parabolic fixed point shown as a
  dot; (b) $\Uc=0.2$, with the hyperbolic fixed point shown as a dot,
  and its associated homoclinic separatrix plotted as a solid line.
  We can clearly see closed trajectories outside the separatrix, near
  the wall.  Once trajectories leave the vicinity of the wall,
  Eq.~\eqref{eq:exactmap} no longer applies and we must look into the
  specific stirring mechanism.}
\label{fig:lines_U}
\end{figure}

To help understand how the velocity~$\Uc$ modifies the wall map, it is
instructive to plot iterates for a specific form of~$\A(\xc)$.  To
match the numerical simulations,~$\A(\xc)$ must be periodic in~$\xc$
and have only two zeros (Fig.~\ref{fig:A_eight}); a simple model is
then to take~$\A(\xc)=-\cos\xc$.  For plotting iterates, it is
preferable to use the exact area-preserving version
of~\eqref{eq:movingwall} (see Appendix~\ref{apx:gen}),
\begin{equation}
  \xcp = \xc + \Uc + \A(\xc)\ycp, \qquad
  \ycp = \yc - \tfrac12\A'(\xc)\ycp^2,
  \label{eq:exactmap}
\end{equation}
which has unit Jacobian.  (Using the exact map guarantees a faithful
representation of closed orbits; but note that~$(\theta,y)$ are only
approximate canonical variables, for small~$y$.)
Figure~\ref{fig:lines_U} shows trajectories for the
map~\eqref{eq:exactmap}.  Once a particle leaves the vicinity of the
wall, its trajectory becomes meaningless, since our
expansions~\eqref{eq:nearwall} and~\eqref{eq:movingwall} are only
valid for small~$\yc$.  However, the cusp structure for~$\Uc=0$ in
Fig.~\ref{fig:lines_U0} is evident and is remarkably similar to the
lower part of Fig.~\ref{fig:fig8exp}, where the unstable separatrix is
located.

Now we analyze the map~\eqref{eq:movingwall} as we did in
Section~\ref{sec:fixed} when the wall was fixed.  Again we look for
fixed points of~\eqref{eq:movingwall}.  All the distinguished
parabolic fixed points on the wall have disappeared, as well as the
two separatrices.  Since~$\A(\xc)$ is continuous, has two zeros,
and~$\A'(\xcsep)>0$, $\A(\xc)$ must have a minimum at~$\xc_1$ and a
maximum at~$\xc_2$, where~$\A'(\xc_{1,2})=0$ and
hence~\hbox{$\ycp(\xc_{1,2},\yc)=\yc$} for all~$\yc$.  Seeking values
for which the along-wall velocity also vanishes, we find there are
fixed points at~$\yc_{1,2} = -\Uc/\A(\xc_{1,2})$, which we require to
be small for our analysis (that is, this defines how slow the rotation
has to be).  Since~$\A(\xc_2)>0$ (maximum) and~$\A(\xc_1)<0$
(minimum), only~$\xc_1$ has~$\yc_1\ge0$.  (Recall that we assume~$\Uc>
0$.)  The other fixed point lies outside our domain.  Hence, we focus
on the unique near-wall fixed point~$(\xc_1,-\Uc/\A(\xc_1))$.  (There
could be others if~$\A(\xc)$ has more extrema, but here we restrict to
the case of two extrema.)

\begin{figure}
  \ifaps{
    \includegraphics[width=.55\textwidth]{fig5}
  }{
    \includegraphics[width=.55\textwidth]{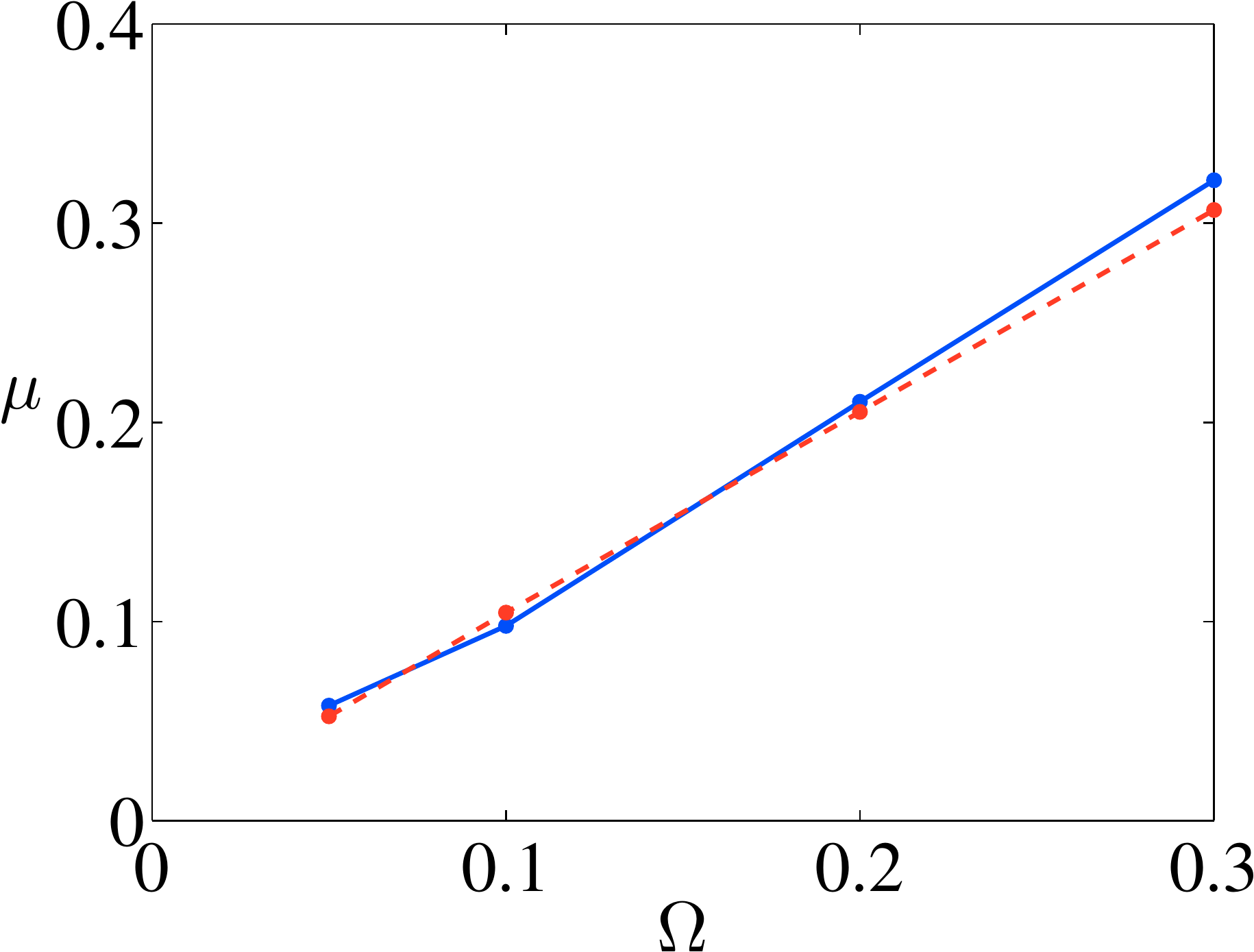}
  }
  \caption{\protect\coloronline The decay rate to the fixed
    point~$\decfp$ measured from simulations (solid line) and from
    Eq.~\eqref{eq:decfp}.  The two converge to each other for
    small~$\Uc$.}
  \label{fig:decfp}
\end{figure}

We look at the linearized dynamics near the fixed point.
Let~$(\xc,\yc) = (\xc_1 + \Xc,-\Uc/\A(\xc_1) + \Yc)$; then
\begin{subequations}
\begin{align}
  \Xcp &= \Xc + \A(\xc_1)\Yc + \Order{\Xc^2,\Yc^2,\Xc\Yc},\\
  \Ycp &= \Yc - \tfrac12\A''(\xc_1)\yc_1^2\,\Xc + \Order{\Xc^2,\Yc^2,\Xc\Yc}.
\end{align}
\end{subequations}%
To leading order in~$\Xc$ and~$\Yc$, the motion near the fixed point
is thus described by
\begin{equation}
  \begin{pmatrix}\Xcp \\ \Ycp\end{pmatrix}
  = 
  \begin{pmatrix}1 & \A(\xc_1) \\ -\tfrac12\A''(\xc_1)\yc_1^2 & 1 \end{pmatrix}
  \begin{pmatrix}\Xc \\ \Yc\end{pmatrix}.
\end{equation}
The matrix in the above equation has eigenvalues
\begin{equation}
  \decfp_\pm = 1 \pm \sqrt{-\tfrac12\A(\xc_1)\A''(\xc_1)}\,\yc_1
  = 1 \pm \sqrt{-\frac{\A''(\xc_1)}{2\A(\xc_1)}}\,\Uc
  \label{eq:ev}
\end{equation}
where the argument in the square root is nonnegative
since~$\A(\xc_1)<0$ and~$\A''(\xc_1)\ge0$.  For~$\A''(\xc_1)>0$
and~$\Uc>0$, this is a hyperbolic fixed point, and the approach along
its stable manifold obeys
\begin{equation}
  \Yc(\time) \sim \Yc(0) \,\exp(-\decfp\,\time/\T)
  \label{eq:asymhyper}
\end{equation}
for~$(\Xc(0),\Yc(0))$ initially on the stable manifold, with decay rate
\begin{equation}
  \decfp = \sqrt{-\tfrac12\A(\xc_1)\A''(\xc_1)}\,\yc_1
  = \sqrt{-\frac{\A''(\xc_1)}{2\A(\xc_1)}}\,\Uc
  \label{eq:decfp}
\end{equation}
to first order in $\Uc$.  Compare this to~\eqref{eq:asympara}: the
approach to the fixed point is now exponential, at a rate proportional
to the speed of rotation of the wall.  In Fig.~\ref{fig:decfp} we show
the results for~$\decfp$ based on Eq.~\eqref{eq:decfp} and as measured
in numerical simulations such as in
Fig.~\ref{fig:eightpoinc_wout=0p2}.  The two agree for small~$\Uc$, as
expected.  In Section~\ref{sec:var}, we will see that the rate of
exponential decay given by~$\decfp$ will dominate if it is slower than
the mixing rate in the bulk.  Otherwise, if~$\decfp$ is large enough,
then the rate of mixing in the bulk dominates.

Figure~\ref{fig:eightpoinc_wout=0p2} shows a Poincar\'e section
for~$\Uc=0.2$.  Very close to the wall (within a distance proportional
to~$\Uc$) trajectories are closed.  A separatrix, consisting of a
homoclinic orbit connecting the hyperbolic fixed point to itself,
isolates the wall region from the bulk.  Now it is the approach to
this separatrix that will limit the decay rate, and as given by
Eq.~\eqref{eq:asymhyper} this approach is exponential.  Since the
iterates of the map are close together on the separatrix, we can find
an equation for the separatrix by writing a stream function
\begin{equation}
  \psi(\xc,\yc) = \tfrac12\A(\xc)\yc^2 + \Uc\yc\,,
  \qquad (\uc,\vc) = (\pd_\yc\psi\,,\,-\pd_\xc\psi)
\end{equation}
for a steady flow corresponding to the map.  The value of the
stream function at the hyperbolic fixed point is
$\psi(\xc_1,-\Uc/\A(\xc_1)) = -\tfrac12\Uc^2/\A(\xc_1)$.  The
separatrix thus satisfies
\begin{equation}
  \tfrac12\A(\xc)\ycsep^2(\xc)
  + \Uc\ycsep(\xc) = -\tfrac12\Uc^2/\A(\xc_1)\,,
\end{equation}
which we can solve for~$\ycsep(\xc)$,
\begin{equation}
  \ycsep(\xc) = \Uc\,\frac{-1 + \sqrt{1 - \A(\xc)/\A(\xc_1)}}{\A(\xc)}
  = \yc_1\,\frac{1 - \sqrt{1 - \A(\xc)/\A(\xc_1)}}{\A(\xc)/\A(\xc_1)}\,.
\end{equation}
where we have chosen the solution of the quadratic that
ensures~$\ycsep(\xc)>0$.  The argument of the square root is always
positive, since~$\xc_1$ is a global minimum and~\hbox{$\A(\xc_1)<0$}.
When~$\xc=\xc_1$ we recover~$\ycsep(\xc_1)=\yc_1$, as required.  The
singularity when~$\A(\xc)=0$ is removable, and we have
then~$\ycsep=\tfrac12\yc_1$.  The distance from the separatrix to the
wall is proportional to~$\Uc$; the point of closest approach to the
wall is at~$\xc_2$ where~$\A(\xc)$ is maximum, and the farthest from
the wall is at~$\xc_1$ where~$\A(\xc)$ is minimum (the hyperbolic
fixed point itself).  Figure~\ref{fig:eightpoinc_wout=0p2} shows the
separatrix and fixed point, computed using the theory in this section
from the observed~$\A(\xc)$ for a moving wall (see
Fig.~\ref{fig:A_eight}, dashed line).  The separatrix clearly divides
the chaotic sea from the wall region.  The cross in
Fig.~\ref{fig:eightpoinc_wout=0p2} is the numerically-computed fixed
point for the map.  The discrepancy between the numerical value and
the theory is due to~$\Uc$ being relatively large in this figure
($\Uc=0.2$), and can be easily accounted for by expanding to next
order in~$\Uc$.

\section{Decay of Variance}
\label{sec:var}

So far in Sections~\ref{sec:fixed} and~\ref{sec:movingwall} we have
shown the following: (i) If the wall is not rotating, then there is a
distinguished parabolic fixed point at the wall with a stable
separatrix; the rate of approach of trajectories along that separatrix
is algebraic.  (ii) If the wall is rotating, the distinguished
parabolic fixed point is destroyed, and instead a hyperbolic fixed
point appears elsewhere, away from the wall; the rate of approach of
trajectories along the stable manifold of that fixed point is
exponential.  We shall now show that the different rates of approach
are reflected in a different time-evolution for the scalar variance.

The key is that the kidney-shaped `mixing pattern,' by which we mean
the shape of a blob that has been advected for several periods of the
protocol approaches the fixed point at a rate~$\gap(\time)$, where
\begin{equation}
  \gap(\time) = \begin{cases}
    2\T/(\A'(\xcsep)\,\time),\qquad &
    \text{fixed wall [Eq.~\eqref{eq:asympara}]};\\
    \gap(0)\,\ee^{-\decfp\time/\T},\qquad &
    \text{moving wall [Eq.~\eqref{eq:asymhyper}]}.
  \end{cases}
  \label{eq:gap}
\end{equation}
(See Fig.~\ref{fig:fig8both} for the definition of~$\gap(\time)$.)
Indeed, a typical fluid particle first approaches the fixed point
before being swept away along the wall or along the separatrix.

Inside the central mixing region, we assume the action of the flow is
that of a simple chaotic mixer.  By this we mean that fluid elements
are stretched, on average, at an exponential rate~$\lyap$.  For
simplicity, we also assume that in the absence of walls the
concentration variance decays at the same `natural' decay
rate~$\lyap$, though in general the two rates can
differ~\cite{Antonsen1996, Balkovsky1999, ThiffeaultAosta2004,
  Meunier2010}.  Hence, after a time~$\time$ a typical blob of initial
size~$\delta$ will have length~$\delta\,\ee^{\lyap\time}$.  However,
because of diffusion, its width will stabilize at an equilibrium
between compression and diffusion at the Batchelor
scale~\cite{Batchelor1959, Balkovsky1999, Villermaux2003,
  ThiffeaultAosta2004}
\begin{equation}
  \lB = \sqrt{\kappa/\lyap},
\end{equation}
where~$\kappa$ is the molecular diffusivity.

At every period, the pattern gets progressively closer to the wall.
Assuming molecular diffusion can be neglected, area preservation
implies that some white fluid must have entered the central mixing
region.  It does so in the form of white strips, clearly visible as
layers inside the pattern of Fig.~\ref{fig:fig8exp}.  In fact, if we
assume that the mixing pattern grows uniformly, we can write the
width~$\Delta(\time)$ of a strip injected at period~$\time/\T$ as
\begin{equation}
  \Delta(\time) = \gap(\time) - \gap(\time+\T) \simeq -\T\,\dotgap(\time)
  \label{eq:Delta}
\end{equation}
where we also assumed that~$\gap(\time)$ changes little at each
period, consistent with experimental observations.  This is positive
since~$\gap$ is a decreasing function of time.

Now, if a white strip is injected at time~$\tau<\time$, how long does
it survive before it is wiped out by diffusion?  The answer is the
solution to the equation
\begin{equation}
  \Delta(\tau)\,\ee^{-\lyap(\time-\tau)} = \lB\,.
  \label{eq:age}
\end{equation}
We interpret this formula as follows: The strip initially has
width~$\Delta(\tau)$ when it is injected; it is then compressed by the
flow in the central mixing region by a
factor~$\ee^{-\lyap(\time-\tau)}$ depending on its age,~$\time-\tau$;
and once it is compressed to the Batchelor length~$\lB$ it quickly
diffuses away.  Thus, we can solve~\eqref{eq:age} to find the age of a
strip when it gets wiped out by diffusion,
\begin{equation}
  \time-\tau = \lyap^{-1}\log(\Delta(\tau)/\lB).
  \label{eq:age2}
\end{equation}

Eventually, at time~$\tB$, a newly-injected filament will have width
equal to the Batchelor length.  This occurs when
\begin{equation}
  \Delta(\tB) = \lB,
  \label{eq:tB}
\end{equation}
which can be solved for~$\tB$ given a form for~$\Delta(\time)$.  After
this time it makes no sense to speak of newly-injected filaments as
`white,' since they are already dominated by diffusion at their birth.
Hence, the description we present here is valid only for times earlier
than~$\tB$, but large enough that the edge of the mixing pattern has
reached the vicinity of the wall.

In the experiment we measure the intensity of pixels in the central
mixing region.  We observe for~\hbox{$1 \ll \time/\T \lesssim \tB/\T$}
that the concentration variance is dominated by the amount of strips
in the central region that are still
white~\cite{Gouillart2007,Gouillart2008}.  Because of area
conservation, the total area of injected white material that is still
visible at time~$\time$ is proportional to
\begin{equation}
  \Aw(\time) = \gap(\tau(\time)) - \gap(\time)
  \label{eq:Aw}
\end{equation}
where we use~\eqref{eq:age2} to solve for~$\tau(\time)$, the injection
time of the oldest strip that is still white at time~$\time$.  Hence,
our goal is to estimate~$\Aw(\time)$ for times~\hbox{$1 \ll \time/\T
  \lesssim \tB/\T$}, since~$\Aw$ is directly proportional to the
concentration variance.  To do this we need~$\tau(\time)$, which
requires specifying~$\Delta(\time)$.  We examine the two possible
forms in Eq.~\eqref{eq:gap} in Sections~\ref{sec:varfixed}
and~\ref{sec:varmoving}.

\subsection{Fixed Wall}
\label{sec:varfixed}

For the fixed wall of Section~\ref{sec:fixed}, we have
from~\eqref{eq:gap}~$\gap(\time) = 2\T/(\A'(\xcsep)\,\time)$.  The
total area of remaining white strips at time~$\time$ as given
by~\eqref{eq:Aw} is proportional to
\begin{equation}
  \Aw(\time) = \frac{2\T}{\A'(\xcsep)\tau} - \frac{2\T}{\A'(\xcsep)\time}
  = \frac{2\T}{\A'(\xcsep)}\,\frac{\time-\tau}{\tau\time}\,.
  \label{eq:Awalg0}
\end{equation}
From~\eqref{eq:gap} and~\eqref{eq:Delta}, the width of injected strips
is~$\Delta(\time) = -\T\dotgap = 2\T^2/(\A'(\xcsep)\time^2)$.
Equation~\eqref{eq:age2} cannot be solved exactly, but
since~$\tau(\time)$ is algebraic the right-hand side
of~\eqref{eq:age2} is not large, implying that~$\time/\tau\simeq 1$
for large~$\time$.  We can thus replace~$\tau$ by~$\time$
in~\eqref{eq:age2} and the denominator of~\eqref{eq:Awalg0}, and
find
\begin{equation}
  \Aw(\time) \simeq \frac{2\T}{\A'(\xcsep)}\,
  \frac{\log(\Delta(\time)/\lB)}{\lyap\,\time^2}\,,\qquad
   1 \ll \time/\T \lesssim \tB/\T.
  \label{eq:Awalg}
\end{equation}
The decay of concentration variance is algebraic ($\sim 1/\time^2$),
with a logarithmic correction.  The form~\eqref{eq:Awalg} has been
verified in experiments and using a simple map
model~\cite{Gouillart2007,Gouillart2008}.

\subsection{Moving Wall}
\label{sec:varmoving}

Now consider the case of a moving wall,
with~$\gap(\time)=\gap(0)\,\ee^{-\decfp\time}$ from
Eq.~(\ref{eq:gap}).  We have~$\Delta(\time)=-\T\dotgap =
\decfp\T\gap(0)\,\ee^{-\decfp\time} = \Delta(0)\,\ee^{-\decfp\time}$.
From~(\ref{eq:tB}), we have~$\tB=\decfp^{-1}\log(\Delta(0)/\lB)$, and
from~(\ref{eq:age2}),
\begin{equation}
  \time - \tau = \frac{\decfp}{\lyap-\decfp}\,(\tB - \time).
  \label{eq:exptau}
\end{equation}
By assumption, $\tau<\time<\tB$, so for consistency we
require~$\decfp<\lyap$, i.e., the rate of approach toward the
hyperbolic fixed point is slower than the natural decay rate of the
chaotic mixer. The area of white material in the mixing region is then
obtained from~\eqref{eq:Aw},
\begin{equation}
  \Aw(\time) = \gap(0)\,\ee^{-\decfp\time}
  \l(\exp\l(\frac{\decfp^2}{\lyap-\decfp}(\tB-\time)\r) - 1\r),
\end{equation}
which in the regime~$\lyap/\decfp\gg 1$ can be approximated by
\begin{equation}
  \Aw(\time) \sim
  \gap(0)\,\lyap^{-1}\decfp^2\,(\tB-\time)\,\ee^{-\decfp\time},
  \qquad \time \lesssim \tB.
  \label{eq:Awexp}
\end{equation}
The decay rate of the `white' area is completely dominated by the rate
of approach to the hyperbolic fixed point.  The central mixing process
is potentially more efficient ($\lyap>\decfp$), but it is starved by
the boundaries.  Hence, the wall slows down mixing, but unlike the
fixed wall (Section~\ref{sec:varfixed}) the decay rate is still
exponential.

If the wall rotates rapidly enough to make $\decfp>\lyap$, then the
wall does not limit the decay rate at all.  Indeed,
for~$\decfp>\lyap$, we have $\time>\tB$ in~\eqref{eq:exptau}, since
newly injected strips reach the Batchelor length~$\lB$ before strips
that were injected previously.  This violates our assumptions, and we
conclude that in that case the white strips can be neglected; the
decay rate of the concentration variance is then given by the natural
decay rate~$\lyap$.

In summary, we expect the variance decay rate to be given by
\begin{equation}
  \decvar = \min(\decfp\,,\lyap),
  \label{eq:decvar}
\end{equation}
that is, the decay rate of the variance is given by either the rate of
approach to the hyperbolic fixed point~$\decfp$, or the `natural'
decay rate~$\lyap$, whichever is slowest. Because the decay rate is
always limited by the natural decay rate~$\lyap$, one should not
rotate the wall faster than is necessary to reach $\decfp \sim \lyap$,
as further increasing the rotation rate only decreases the size of the
chaotic region, without any improvement to the mixing rate. We will
see in the next section how this compares to simulations.

\section{Simulation Results and Discussion}
\label{sec:simulation}

\begin{figure}
  \ifaps{
    \includegraphics[width=.4\textwidth]{fig6}
  }{
    \includegraphics[width=.4\textwidth]{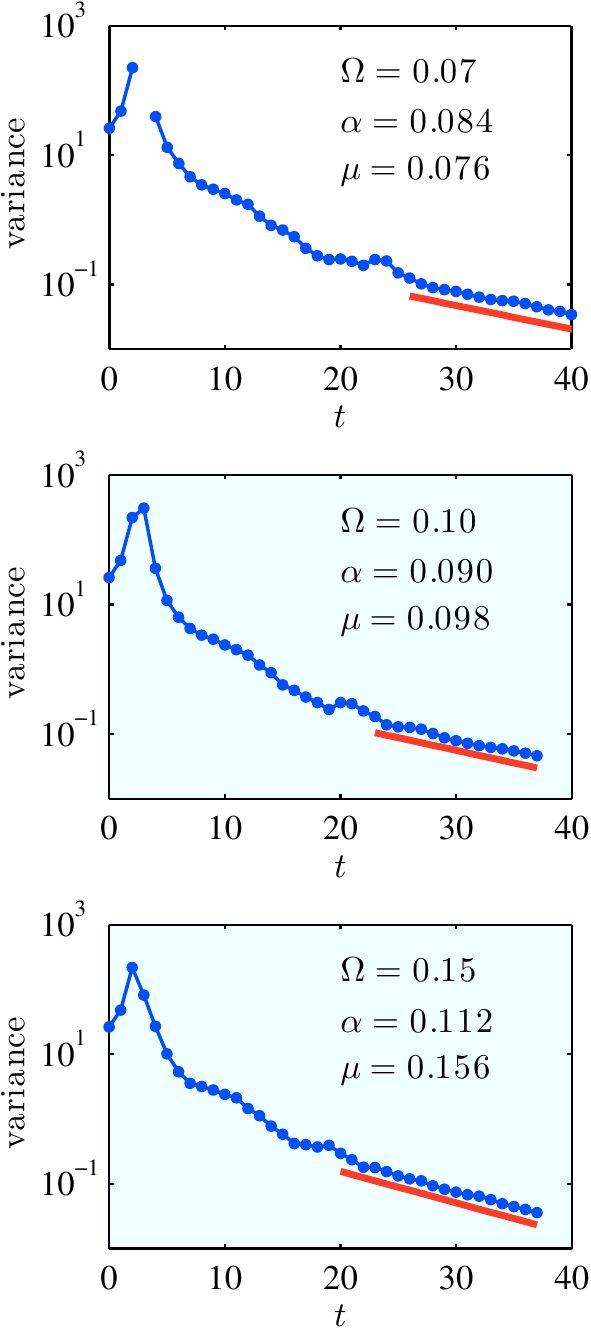}
  }
  \caption{\protect\coloronline Evolution of variance as a function of
    period, for different values of the rotation rate~$\Uc$.  $\alpha$
    is the fitted asymptotic decay rate (solid line).  For
    small~$\Uc$, $\alpha$ tends to~$\decfp$ --- the decay rate onto
    the hyperbolic fixed point (see Eq.~\eqref{eq:decfp}).  (The
    variance is measured over a finite subregion of the domain, which
    means it can increase at early times; the break in the top figure
    is due to an accidental absence of any particles in our
    measurement region, before particles get mixed.)}
  \label{fig:var}
\end{figure}

The case of an algebraic decay was well-documented
in~\cite{Gouillart2007,Gouillart2008,Gouillart2009}, so we focus here on the
exponential decay for a moving wall.  Figure~\ref{fig:var} shows the
evolution of variance for several angular rotation rates~$\Uc$.
\begin{figure}
  \ifaps{
    \includegraphics[width=.55\textwidth]{fig7}
  }{
    \includegraphics[width=.55\textwidth]{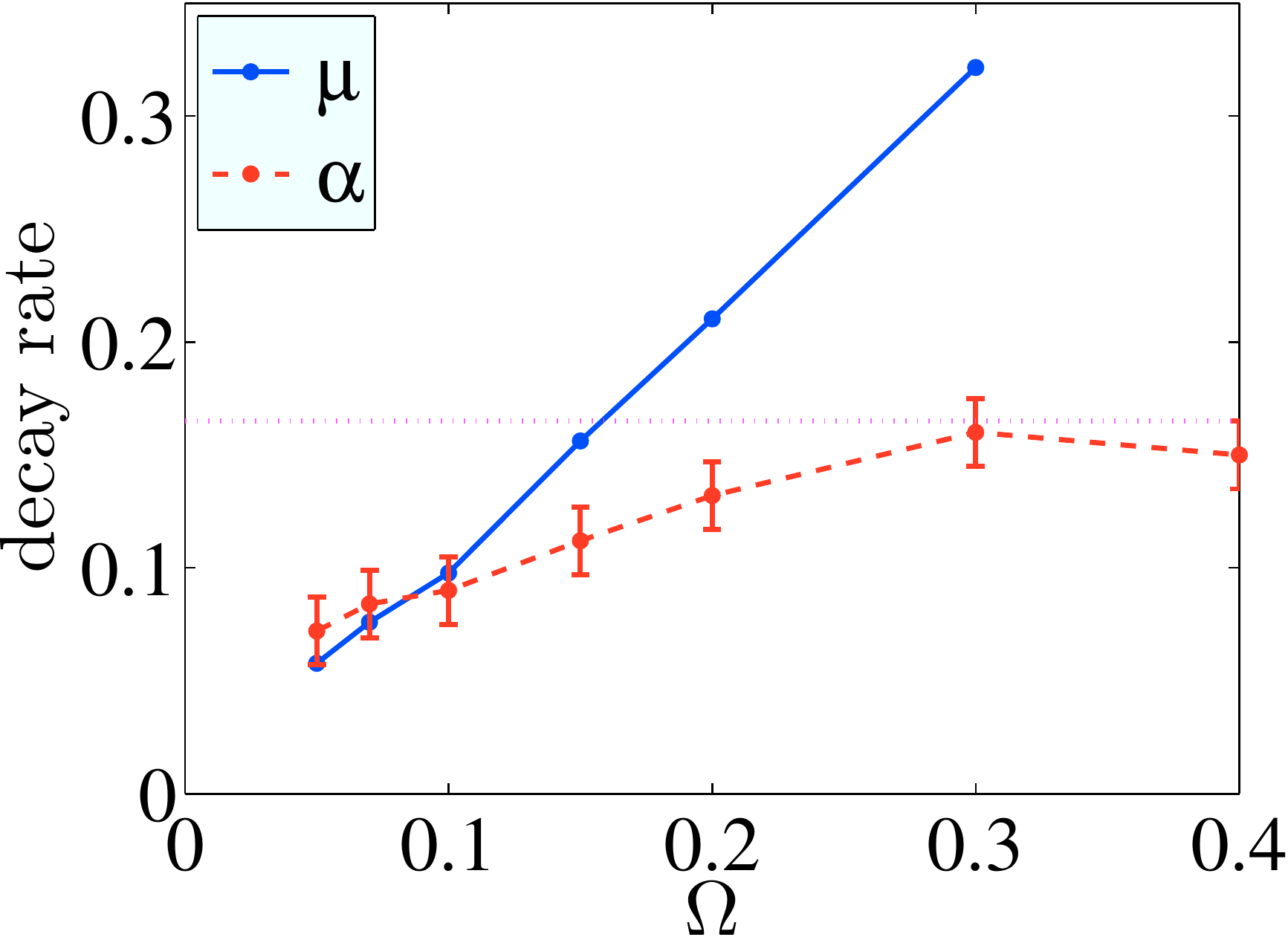}
  }
  \caption{\protect\coloronline Rate of approach to the fixed point
    $\decfp$ (solid line) and rate of decay of variance $\alpha$
    (dashed line).}
  \label{fig:dec}
\end{figure}%
The variance was obtained by evolving a large number (of order~4
million) of particles to model the concentration field of a passive
scalar.  This coarse method does not allow us to evolve the
concentration field for a long time before particle statistics become
inadequate, but we can conclude that the system enters an exponential
regime for~$\time/\T \gtrsim 25$, with a rate of decay~$\alpha$.

Figure~\ref{fig:dec} shows the numerically-measured decay
rate~$\decfp$ to the hyperbolic fixed point near the wall (solid
line), and the decay rate of the concentration variance (dashed line
with error bars).  For small~$\Uc$, the exponential rate of decay is
approximately equal to~$\decfp$; for larger~$\Uc$ it is less
than~$\decfp$, since the decay rate is no longer dominated by the
approach to the hyperbolic fixed point.  In fact, as predicted by the
theory, as the rotation rate is increased the decay rate tends to its
``natural'' value, approximated here by the dotted line.  For small
rotation rates, the agreement between~$\decfp$ and~$\alpha$ cannot be
called convincing, but is at least consistent.  This is partly due to
difficulties in measuring the decay rate accurately, as reflected by
the large error bars on the dashed line.  Note also that the effect is
rather small here, so for this particular system the decay rate is not
greatly limited by the walls.  Nevertheless, we argue that the effect
is there, and that in other situations this might be a more
significant factor.  For instance, it is known that the motion of
walls can enhance heat transfer~\cite{ElOmari2009,ElOmari2010}.

\begin{figure}
\subfigure[]{
  \ifaps{
    \includegraphics[width=.4\textwidth]{fig8a}
  }{
    \includegraphics[width=.4\textwidth]{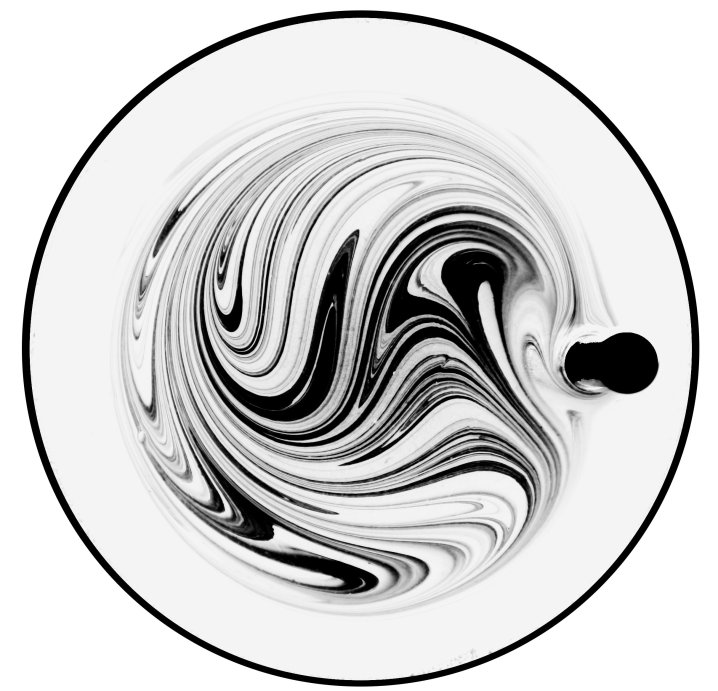}
  }
  \label{fig:trm1}
}\hspace{2em}%
\subfigure[]{
  \ifaps{
    \includegraphics[width=.384\textwidth]{fig8b}
  }{
    \includegraphics[width=.384\textwidth]{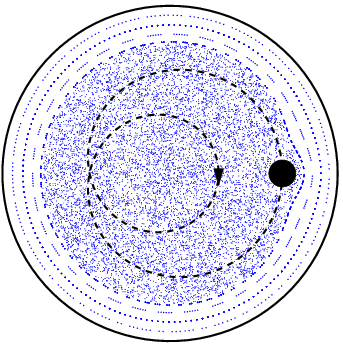}
  }
  \label{fig:trmpoinc}
}
\caption{The `epitrochoid' stirring protocol.  (a) Experiment; (b)
  Poincar\'e section, also showing the rod's trajectory.  Notice the
  presence of closed orbits near the wall, even though the wall is
  fixed.}
\label{fig:trm}
\end{figure}

\begin{figure}
  \ifaps{
    \includegraphics[width=.5\textwidth]{fig9}
  }{
    \includegraphics[width=.5\textwidth]{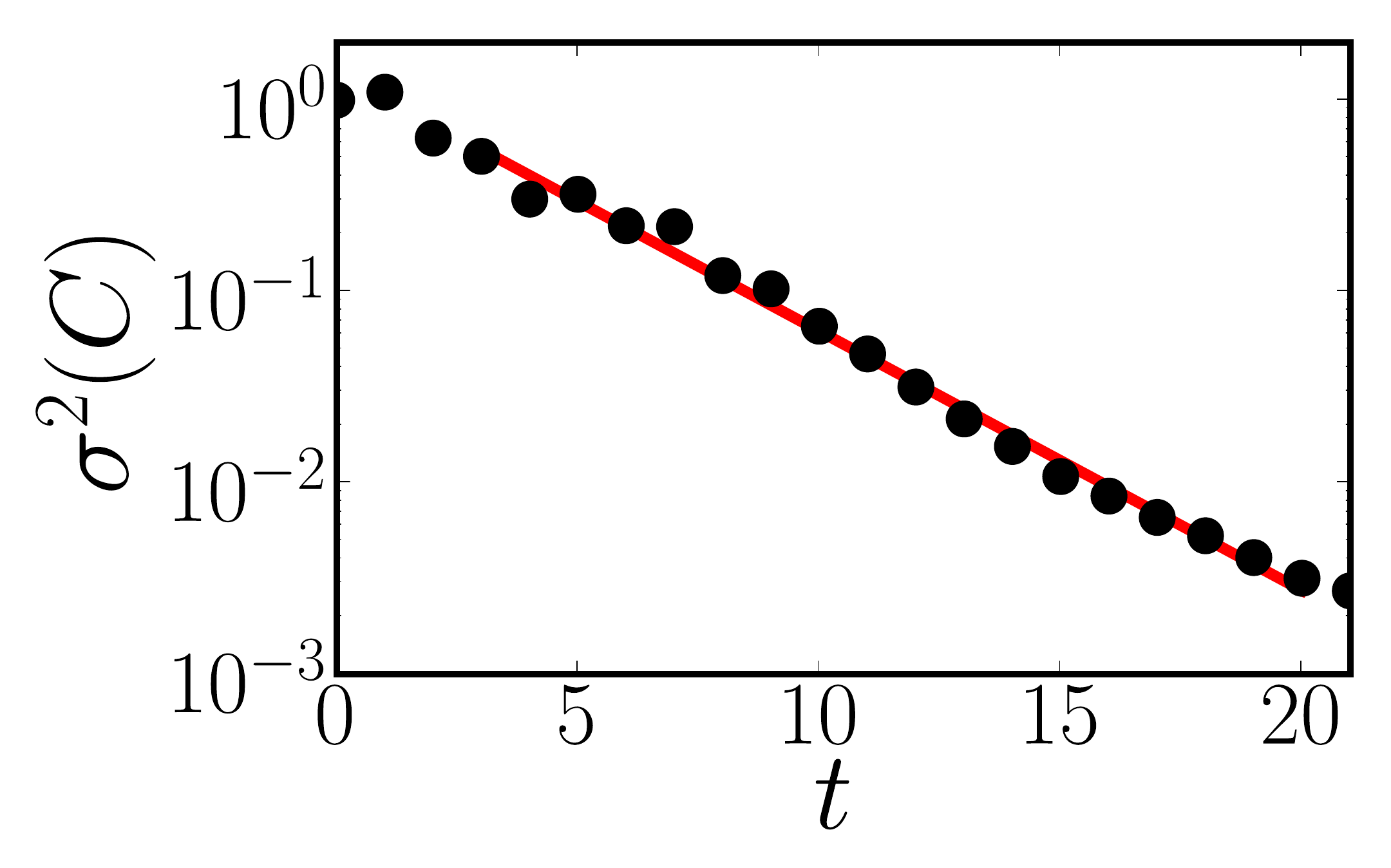}
  }
  \caption{Decay of concentration variance for the epitrochoid
    stirring protocol (Fig.~\ref{fig:trm}), exhibiting an exponential
    decay.}
\label{fig:var_trm_pres}
\end{figure}

Evidence that this type of effect could be important is provided by
Figs.~\ref{fig:trm} and~\ref{fig:var_trm_pres}.
They show the results of an experiment where the rod is moved along a
``double-loop'' rather than the figure-eight (see
also~\cite{Thiffeault2008b} for a discussion of this system).  The
wall is fixed. Many industrial mixers, for example the wide class of
\emph{planetary mixers}~\cite{MattFinn2001, Clifford2004}, impose such
a multi-loop motion to paddles or stirrers. The double-loop motion,
unlike the figure-eight, induces a net angular displacement of
particles near the wall, leading to closed orbits
(Fig.~\ref{fig:trmpoinc}).  These closed orbits provide a similar
effect to a moving wall: they prevent separatrices from being
connected to the wall. Hence, the decay of concentration variance
should be exponential, which is confirmed by the experimental results
in Fig.~\ref{fig:var_trm_pres}.  The analysis of such a system,
however, is significantly harder than the moving wall case presented
here.  (Note that these two cases are not simply related by changing
to a rotating frame: in general the rotation frequency of the wall and
the period of the rods are not rationally related, so a fixed frame
does not exist.)  A detailed study of this case along the lines
described here would certainly shed some light on the transition from
algebraic to exponential decay, and how system parameters could be
tuned for improved mixing efficiency.

Another important generalization is to extend the analysis to three
dimensions, where even steady flows can exhibit exponential decay. This
is particularly important for microfluidics
applications~\cite{Stroock2002, Villermaux2008}, where the walls play an
important role. The same analysis of separatrices at the walls should
carry through, albeit with a richer range of possibilities that is
inherent to three dimensions.

\subsection*{Acknowledgments}

The authors thank Matthew D. Finn for the use of his computer code for
simulating viscous flows, and an anonymous referee for suggesting the
area-preserving form of the map in the appendix.  J-LT was also
partially supported by the Division of Mathematical Sciences of the US
National Science Foundation, under grant DMS-0806821.

\appendix

\section{Area-preserving Maps}
\label{apx:gen}

The near-wall maps~\eqref{eq:nearwall} and~\eqref{eq:movingwall} were
derived by assuming a simple Taylor expansion in the vicinity of the
wall, and applying boundary conditions.  However, maps suffer from
some drawbacks: they do not exactly preserve area, and they do not
satisfy a natural time-reversibility formula.  Though these drawback
do not quantitatively modify the predictions of the paper, for
completeness in this appendix we remedy the situtation.

We wish to derive a symplectic (area-preserving) map for the flow
having Hamiltonian (streamfunction)
\begin{equation}
  \Ham(\xc,\yc,\time) = \omega\yc + \tfrac12\aaa(\theta)\yc^2
  \sum_{k=-\infty}^{\infty}\delta(\time - k\T).
  \label{eq:theHam}
\end{equation}
In this flow the fluid near the wall is rotated and `kicked' at each
period.  The second term in~\eqref{eq:theHam} is a perturbation of the
integrable first term.  Following~\citet{Abdullaev}, the simplest way
to obtain an area-preserving map from~\eqref{eq:theHam} is to
introduce the generating function
\begin{equation}
  \Fgen(\xc,\ycp) = (\xc+\Uc)\ycp + \tfrac12\A(\xcp)\ycp^2,
\end{equation}
with~$\A(\xc)=\aaa(\xc)\T$ and~$\Uc=\omega\T$.  With this type of
generating function the mapping~$(\xc,\yc) \rightarrow (\xcp,\ycp)$ is
defined implicitly via
\begin{equation}
  \xcp = \frac{\pd\Fgen(\xc,\ycp)}{\pd\ycp},\qquad
  \yc = \frac{\pd\Fgen(\xc,\ycp)}{\pd\xc},
  \label{eq:mapfromgen}
\end{equation}
which gives the map
\begin{equation}
  \xcp = \xc + \Uc + \A(\xc)\ycp,\qquad
  \ycp = \yc - \tfrac12\A'(\xc)\ycp^2.
  \label{eq:movingwallAPasym}
\end{equation}
Note that this map is the same as~\eqref{eq:movingwall} to leading
order in~$\yc$.  However, unlike~\eqref{eq:movingwall} it exactly
preserves area, by construction.  One can see this directly from the
generating function~$\Fgen(\xc,\ycp) = \Fgen(\xc,\ycp(\xc,\yc))$: from
the second equation of~\eqref{eq:mapfromgen}, we have
\begin{equation}
  \frac{\pd\yc}{\pd\xc} = 0
  = \Fgen_{\xc\xc} + \Fgen_{\xc\ycp}\frac{\pd\ycp}{\pd\xc}
  \quad\Longleftrightarrow\quad
  \frac{\pd\ycp}{\pd\xc} = -\frac{\Fgen_{\xc\xc}}{\Fgen_{\xc\ycp}},
\end{equation}
as well as
\begin{equation}
  \frac{\pd\yc}{\pd\yc} = 1
  = \Fgen_{\xc\ycp}\frac{\pd\ycp}{\pd\yc}
  \quad\Longleftrightarrow\quad
  \frac{\pd\ycp}{\pd\yc} = \frac{1}{\Fgen_{\xc\ycp}}.
\end{equation}
From the first equation of~\eqref{eq:mapfromgen}, we have
\begin{equation}
  \frac{\pd\xcp}{\pd\xc}
  = \Fgen_{\ycp\xc}\frac{\pd\xc}{\pd\xc} + \Fgen_{\ycp\ycp}\frac{\pd\ycp}{\pd\xc}
  = \Fgen_{\xc\ycp} - \frac{\Fgen_{\xc\xc}\Fgen_{\ycp\ycp}}{\Fgen_{\xc\ycp}},
\end{equation}
as well as
\begin{equation}
  \frac{\pd\xcp}{\pd\yc}
  = \Fgen_{\ycp\ycp}\frac{\pd\ycp}{\pd\yc}
  = \frac{\Fgen_{\ycp\ycp}}{\Fgen_{\xc\ycp}}.
\end{equation}
Assembling all this, we find the linearization matrix
\begin{equation}
  \frac{\pd(\xcp,\ycp)}{\pd(\xc,\yc)} =
  \frac{1}{\Fgen_{\xc\ycp}}
  \begin{pmatrix}
    \Fgen_{\xc\ycp}^2 - \Fgen_{\xc\xc}\Fgen_{\ycp\ycp} &
    \Fgen_{\ycp\ycp} \\
    -\Fgen_{\xc\xc} & 1
  \end{pmatrix}
\end{equation}
which always has unit determinant, as required for area preservation.
The area-preserving map~\eqref{eq:movingwallAPasym} can be written in
explicit form by solving for~$\ycp$ in the second equation,
\begin{equation}
  \ycp = \l(\sqrt{1 + 2\A'(\theta)\yc} - 1\r)/\A'(\theta)\,,
\end{equation}
where the sign of the quadratic solution was chosen so that the map
reduces to~\eqref{eq:movingwall} for small~$\yc$.

The map~\eqref{eq:movingwallAPasym} solves one of the issues mentioned
at the outset: it exactly preserves area.  However, it fails to
resolve the second issue, time-reversal symmetry.  Indeed,
interchanging~$(\xc,\yc)$ with~$(\xcp,\ycp)$ and letting~$\T
\rightarrow -\T$ in~\eqref{eq:movingwallAPasym} leads to a different
form for the map, which should not be the case for a map arising from
a Hamiltonian such as~\eqref{eq:theHam}.  To fix this is more
complicated and requires the introduction of auxiliary
variables~$(\vartheta,\Upsilon)$~\cite{Abdullaev}.  We quote the form
of the resulting map here for completeness:
\begin{subequations}
\begin{gather}
  \Upsilon = \yc - \tfrac14\A'(\xc)\Upsilon^2, \qquad
  \vartheta = \theta + \tfrac12\A(\theta)\Upsilon, \\
  \xcp = \bar\vartheta + \tfrac12\A(\xcp)\Upsilon, \qquad
  \ycp = \Upsilon - \tfrac14\A'(\xcp)\Upsilon^2, \qquad
  \bar\vartheta = \vartheta + \Uc.
  \label{eq:movingwallAPsym2}%
\end{gather}%
\label{eq:movingwallAPsym}%
\end{subequations}
This map is both area-preserving and invariant under interchange of
the barred variables together with~$\T \rightarrow -\T$ (the latter is
equivalent to~$\A \rightarrow -\A$ and~$\Uc \rightarrow -\Uc$).  We
can eliminate the~$\Upsilon$ and~$\vartheta$ auxiliary variables
in~\eqref{eq:movingwallAPsym} by solving for~$\Upsilon$,
\begin{equation}
  \Upsilon = 2\l(\sqrt{1 + \A'(\theta)\yc} - 1\r)/\A'(\theta)\,,
\end{equation}
but the first equation in~\eqref{eq:movingwallAPsym2} must still be
solved for~$\xcp$, usually numerically.

\bibliographystyle{jlt}
\bibliography{bib/journals_abbrev,bib/articles}

\begin{thebibliography}{46}
\newcommand{\enquote}[1]{`#1'}
\providecommand{\natexlab}[1]{#1}
\providecommand{\url}[1]{\texttt{#1}}
\providecommand{\urlprefix}{URL }
\providecommand{\bibinfo}[2]{#2}
\providecommand{\eprint}[2][]{\url{#2}}

\bibitem[{Aref(1984)}]{Aref1984}
\bibinfo{author}{H.~Aref}, \enquote{\bibinfo{title}{Stirring by chaotic
  advection},} \emph{\bibinfo{journal}{J. Fluid Mech.}}
  \textbf{\bibinfo{volume}{143}}, \bibinfo{pages}{1--21}
  (\bibinfo{year}{1984}).

\bibitem[{Ottino(1989)}]{Ottino}
\bibinfo{author}{J.~M. Ottino}, \emph{\bibinfo{title}{The Kinematics of Mixing:
  Stretching, Chaos, and Transport}} (\bibinfo{publisher}{Cambridge University
  Press}, \bibinfo{address}{Cambridge, U.K.}, \bibinfo{year}{1989}).

\bibitem[{Gouillart \emph{et~al.}(2007)Gouillart, Kuncio, Dauchot, Dubrulle,
  Roux, and Thiffeault}]{Gouillart2007}
\bibinfo{author}{E.~Gouillart}, \bibinfo{author}{N.~Kuncio},
  \bibinfo{author}{O.~Dauchot}, \bibinfo{author}{B.~Dubrulle},
  \bibinfo{author}{S.~Roux}, and \bibinfo{author}{J.-L. Thiffeault},
  \enquote{\bibinfo{title}{Walls inhibit chaotic mixing},}
  \emph{\bibinfo{journal}{Phys. Rev. Lett.}} \textbf{\bibinfo{volume}{99}},
  \bibinfo{pages}{114501} (\bibinfo{year}{2007}).

\bibitem[{Gouillart \emph{et~al.}(2008)Gouillart, Dauchot, Dubrulle, Roux, and
  Thiffeault}]{Gouillart2008}
\bibinfo{author}{E.~Gouillart}, \bibinfo{author}{O.~Dauchot},
  \bibinfo{author}{B.~Dubrulle}, \bibinfo{author}{S.~Roux}, and
  \bibinfo{author}{J.-L. Thiffeault}, \enquote{\bibinfo{title}{Slow decay of
  concentration variance due to no-slip walls in chaotic mixing},}
  \emph{\bibinfo{journal}{Phys. Rev. E}} \textbf{\bibinfo{volume}{78}},
  \bibinfo{pages}{026211} (\bibinfo{year}{2008}),
  \eprint{http://arxiv.org/abs/0803.0709}.

\bibitem[{Danckwerts(1952)}]{Danckwerts1952}
\bibinfo{author}{P.~V. Danckwerts}, \enquote{\bibinfo{title}{The definition and
  measurement of some characteristics of mixtures},}
  \emph{\bibinfo{journal}{Appl. Sci. Res. A}} \textbf{\bibinfo{volume}{A3}},
  \bibinfo{pages}{279--296} (\bibinfo{year}{1952}).

\bibitem[{Rothstein \emph{et~al.}(1999)Rothstein, Henry, and
  Gollub}]{Rothstein1999}
\bibinfo{author}{D.~Rothstein}, \bibinfo{author}{E.~Henry}, and
  \bibinfo{author}{J.~P. Gollub}, \enquote{\bibinfo{title}{Persistent patterns
  in transient chaotic fluid mixing},} \emph{\bibinfo{journal}{Nature}}
  \textbf{\bibinfo{volume}{401}}~(\bibinfo{number}{6755}),
  \bibinfo{pages}{770--772} (\bibinfo{year}{1999}).

\bibitem[{Jullien \emph{et~al.}(2000)Jullien, Castiglione, and
  Tabeling}]{Jullien2000}
\bibinfo{author}{M.-C. Jullien}, \bibinfo{author}{P.~Castiglione}, and
  \bibinfo{author}{P.~Tabeling}, \enquote{\bibinfo{title}{Experimental
  observation of batchelor dispersion of passive tracers},}
  \emph{\bibinfo{journal}{Phys. Rev. Lett.}} \textbf{\bibinfo{volume}{85}},
  \bibinfo{pages}{3636} (\bibinfo{year}{2000}).

\bibitem[{Fereday \emph{et~al.}(2002)Fereday, Haynes, Wonhas, and
  Vassilicos}]{Fereday2002}
\bibinfo{author}{D.~R. Fereday}, \bibinfo{author}{P.~H. Haynes},
  \bibinfo{author}{A.~Wonhas}, and \bibinfo{author}{J.~C. Vassilicos},
  \enquote{\bibinfo{title}{Scalar variance decay in chaotic advection and
  {B}atchelor-regime turbulence},} \emph{\bibinfo{journal}{Phys. Rev. E}}
  \textbf{\bibinfo{volume}{65}}~(\bibinfo{number}{3}),
  \bibinfo{pages}{035301(R)} (\bibinfo{year}{2002}).

\bibitem[{Sukhatme and Pierrehumbert(2002)}]{Sukhatme2002}
\bibinfo{author}{J.~Sukhatme} and \bibinfo{author}{R.~T. Pierrehumbert},
  \enquote{\bibinfo{title}{Decay of passive scalars under the action of single
  scale smooth velocity fields in bounded two-dimensional domains: From
  non-self-similar probability distribution functions to self-similar
  eigenmodes},} \emph{\bibinfo{journal}{Phys. Rev. E}}
  \textbf{\bibinfo{volume}{66}}, \bibinfo{pages}{056302}
  (\bibinfo{year}{2002}).

\bibitem[{Voth \emph{et~al.}(2003)Voth, Saint, Dobler, and Gollub}]{Voth2003}
\bibinfo{author}{G.~A. Voth}, \bibinfo{author}{T.~C. Saint},
  \bibinfo{author}{G.~Dobler}, and \bibinfo{author}{J.~P. Gollub},
  \enquote{\bibinfo{title}{Mixing rates and symmetry breaking in
  two-dimensional chaotic flow},} \emph{\bibinfo{journal}{Phys. Fluids}}
  \textbf{\bibinfo{volume}{15}}~(\bibinfo{number}{9}),
  \bibinfo{pages}{2560--2566} (\bibinfo{year}{2003}).

\bibitem[{Villermaux and Duplat(2003)}]{Villermaux2003}
\bibinfo{author}{E.~Villermaux} and \bibinfo{author}{J.~Duplat},
  \enquote{\bibinfo{title}{Mixing as an aggregation process},}
  \emph{\bibinfo{journal}{Phys. Rev. Lett.}} \textbf{\bibinfo{volume}{91}},
  \bibinfo{pages}{184501} (\bibinfo{year}{2003}).

\bibitem[{Thiffeault and Childress(2003)}]{Thiffeault2003d}
\bibinfo{author}{J.-L. Thiffeault} and \bibinfo{author}{S.~Childress},
  \enquote{\bibinfo{title}{Chaotic mixing in a torus map},}
  \emph{\bibinfo{journal}{Chaos}}
  \textbf{\bibinfo{volume}{13}}~(\bibinfo{number}{2}),
  \bibinfo{pages}{502--507} (\bibinfo{year}{2003}).

\bibitem[{Zeldovich \emph{et~al.}(1984)Zeldovich, Ruzmaikin, Molchanov, and
  Sokoloff}]{Zeldovich1984}
\bibinfo{author}{Y.~B. Zeldovich}, \bibinfo{author}{A.~A. Ruzmaikin},
  \bibinfo{author}{S.~A. Molchanov}, and \bibinfo{author}{D.~D. Sokoloff},
  \enquote{\bibinfo{title}{Kinematic dynamo problem in a linear velocity
  field},} \emph{\bibinfo{journal}{J. Fluid Mech.}}
  \textbf{\bibinfo{volume}{144}}, \bibinfo{pages}{1--11}
  (\bibinfo{year}{1984}).

\bibitem[{Shraiman and Siggia(1994)}]{Shraiman1994}
\bibinfo{author}{B.~I. Shraiman} and \bibinfo{author}{E.~D. Siggia},
  \enquote{\bibinfo{title}{{L}agrangian path integrals and fluctuations in
  random flow},} \emph{\bibinfo{journal}{Phys. Rev. E}}
  \textbf{\bibinfo{volume}{49}}~(\bibinfo{number}{4}),
  \bibinfo{pages}{2912--2927} (\bibinfo{year}{1994}).

\bibitem[{{Antonsen, Jr.} \emph{et~al.}(1996){Antonsen, Jr.}, Fan, Ott, and
  Garcia-Lopez}]{Antonsen1996}
\bibinfo{author}{T.~M. {Antonsen, Jr.}}, \bibinfo{author}{Z.~Fan},
  \bibinfo{author}{E.~Ott}, and \bibinfo{author}{E.~Garcia-Lopez},
  \enquote{\bibinfo{title}{The role of chaotic orbits in the determination of
  power spectra},} \emph{\bibinfo{journal}{Phys. Fluids}}
  \textbf{\bibinfo{volume}{8}}~(\bibinfo{number}{11}),
  \bibinfo{pages}{3094--3104} (\bibinfo{year}{1996}).

\bibitem[{Balkovsky and Fouxon(1999)}]{Balkovsky1999}
\bibinfo{author}{E.~Balkovsky} and \bibinfo{author}{A.~Fouxon},
  \enquote{\bibinfo{title}{Universal long-time properties of {L}agrangian
  statistics in the {B}atchelor regime and their application to the passive
  scalar problem},} \emph{\bibinfo{journal}{Phys. Rev. E}}
  \textbf{\bibinfo{volume}{60}}~(\bibinfo{number}{4}),
  \bibinfo{pages}{4164--4174} (\bibinfo{year}{1999}).

\bibitem[{Thiffeault(2008)}]{ThiffeaultAosta2004}
\bibinfo{author}{J.-L. Thiffeault}, \enquote{\bibinfo{title}{Scalar decay in
  chaotic mixing},} in \bibinfo{editor}{J.~B. Weiss} and
  \bibinfo{editor}{A.~Provenzale}, eds., \emph{\bibinfo{booktitle}{Transport
  and Mixing in Geophysical Flows}}, \emph{\bibinfo{series}{Lecture Notes in
  Physics}}, volume \bibinfo{volume}{744}, \bibinfo{pages}{3--35}
  (\bibinfo{publisher}{Springer}, \bibinfo{address}{Berlin},
  \bibinfo{year}{2008}), \eprint{arXiv:nlin/0502011}.

\bibitem[{Haynes and Vanneste(2005)}]{Haynes2005}
\bibinfo{author}{P.~H. Haynes} and \bibinfo{author}{J.~Vanneste},
  \enquote{\bibinfo{title}{What controls the decay of passive scalars in smooth
  flows?}} \emph{\bibinfo{journal}{Phys. Fluids}}
  \textbf{\bibinfo{volume}{17}}, \bibinfo{pages}{097103}
  (\bibinfo{year}{2005}).

\bibitem[{Meunier and Villermaux(2010)}]{Meunier2010}
\bibinfo{author}{P.~Meunier} and \bibinfo{author}{E.~Villermaux},
  \enquote{\bibinfo{title}{The diffusive strip method for scalar mixing in two
  dimensions},} \emph{\bibinfo{journal}{J. Fluid Mech.}}
  \textbf{\bibinfo{volume}{662}}, \bibinfo{pages}{134--172}
  (\bibinfo{year}{2010}).

\bibitem[{Batchelor(1959)}]{Batchelor1959}
\bibinfo{author}{G.~K. Batchelor}, \enquote{\bibinfo{title}{Small-scale
  variation of convected quantities like temperature in turbulent fluid: {P}art
  1. {G}eneral discussion and the case of small conductivity},}
  \emph{\bibinfo{journal}{J. Fluid Mech.}} \textbf{\bibinfo{volume}{5}},
  \bibinfo{pages}{113--133} (\bibinfo{year}{1959}).

\bibitem[{Pierrehumbert(1994)}]{Pierrehumbert1994}
\bibinfo{author}{R.~T. Pierrehumbert}, \enquote{\bibinfo{title}{Tracer
  microstructure in the large-eddy dominated regime},}
  \emph{\bibinfo{journal}{Chaos Solitons Fractals}}
  \textbf{\bibinfo{volume}{4}}~(\bibinfo{number}{6}),
  \bibinfo{pages}{1091--1110} (\bibinfo{year}{1994}).

\bibitem[{Wonhas and Vassilicos(2002)}]{Wonhas2002}
\bibinfo{author}{A.~Wonhas} and \bibinfo{author}{J.~C. Vassilicos},
  \enquote{\bibinfo{title}{Mixing in fully chaotic flows},}
  \emph{\bibinfo{journal}{Phys. Rev. E}} \textbf{\bibinfo{volume}{66}},
  \bibinfo{pages}{051205} (\bibinfo{year}{2002}).

\bibitem[{Pikovsky and Popovych(2003)}]{Pikovsky2003}
\bibinfo{author}{A.~Pikovsky} and \bibinfo{author}{O.~Popovych},
  \enquote{\bibinfo{title}{Persistent patterns in deterministic mixing flows},}
  \emph{\bibinfo{journal}{Europhys. Lett.}} \textbf{\bibinfo{volume}{61}},
  \bibinfo{pages}{625--631} (\bibinfo{year}{2003}).

\bibitem[{Fereday and Haynes(2004)}]{Fereday2004}
\bibinfo{author}{D.~R. Fereday} and \bibinfo{author}{P.~H. Haynes},
  \enquote{\bibinfo{title}{Scalar decay in two-dimensional chaotic advection
  and {B}atchelor-regime turbulence},} \emph{\bibinfo{journal}{Phys. Fluids}}
  \textbf{\bibinfo{volume}{16}}~(\bibinfo{number}{12}),
  \bibinfo{pages}{4359--4370} (\bibinfo{year}{2004}).

\bibitem[{Jones and Young(1994)}]{Jones1994}
\bibinfo{author}{S.~W. Jones} and \bibinfo{author}{W.~R. Young},
  \enquote{\bibinfo{title}{Shear dispersion and anomalous diffusion by chaotic
  advection},} \emph{\bibinfo{journal}{J. Fluid Mech.}}
  \textbf{\bibinfo{volume}{280}}, \bibinfo{pages}{149--172}
  (\bibinfo{year}{1994}).

\bibitem[{Chertkov and Lebedev(2003)}]{Chertkov2003}
\bibinfo{author}{M.~Chertkov} and \bibinfo{author}{V.~Lebedev},
  \enquote{\bibinfo{title}{Decay of scalar turbulence revisited},}
  \emph{\bibinfo{journal}{Phys. Rev. Lett.}}
  \textbf{\bibinfo{volume}{90}}~(\bibinfo{number}{3}), \bibinfo{pages}{034501}
  (\bibinfo{year}{2003}).

\bibitem[{Lebedev and Turitsyn(2004)}]{Lebedev2004}
\bibinfo{author}{V.~V. Lebedev} and \bibinfo{author}{K.~S. Turitsyn},
  \enquote{\bibinfo{title}{Passive scalar evolution in peripheral regions},}
  \emph{\bibinfo{journal}{Phys. Rev. E}} \textbf{\bibinfo{volume}{69}},
  \bibinfo{pages}{036301} (\bibinfo{year}{2004}).

\bibitem[{Schekochihin \emph{et~al.}(2004)Schekochihin, Haynes, and
  Cowley}]{Schekochihin2004}
\bibinfo{author}{A.~A. Schekochihin}, \bibinfo{author}{P.~H. Haynes}, and
  \bibinfo{author}{S.~C. Cowley}, \enquote{\bibinfo{title}{Diffusion of passive
  scalar in a finite-scale random flow},} \emph{\bibinfo{journal}{Phys. Rev.
  E}} \textbf{\bibinfo{volume}{70}}, \bibinfo{pages}{046304}
  (\bibinfo{year}{2004}).

\bibitem[{Salman and Haynes(2007)}]{Salman2007}
\bibinfo{author}{H.~Salman} and \bibinfo{author}{P.~H. Haynes},
  \enquote{\bibinfo{title}{A numerical study of passive scalar evolution in
  peripheral regions},} \emph{\bibinfo{journal}{Phys. Fluids}}
  \textbf{\bibinfo{volume}{19}}, \bibinfo{pages}{067101}
  (\bibinfo{year}{2007}).

\bibitem[{Popovych \emph{et~al.}(2007)Popovych, Pikovsky, and
  Eckhardt}]{Popovych2007}
\bibinfo{author}{O.~V. Popovych}, \bibinfo{author}{A.~Pikovsky}, and
  \bibinfo{author}{B.~Eckhardt}, \enquote{\bibinfo{title}{Abnormal mixing of
  passive scalars in chaotic flows},} \emph{\bibinfo{journal}{Phys. Rev. E}}
  \textbf{\bibinfo{volume}{75}}, \bibinfo{pages}{036308}
  (\bibinfo{year}{2007}).

\bibitem[{Chernykh and Lebedev(2008)}]{Chernykh2008}
\bibinfo{author}{A.~Chernykh} and \bibinfo{author}{V.~Lebedev},
  \enquote{\bibinfo{title}{Passive scalar structures in peripheral regions of
  random flows},} \emph{\bibinfo{journal}{{JETP} Lett.}}
  \textbf{\bibinfo{volume}{87}}~(\bibinfo{number}{12}),
  \bibinfo{pages}{682--686} (\bibinfo{year}{2008}).

\bibitem[{Mackay(2008)}]{Mackay_CCT2007}
\bibinfo{author}{R.~S. Mackay}, \enquote{\bibinfo{title}{A steady mixing flow
  with no-slip boundaries},} in \bibinfo{editor}{C.~Chandre},
  \bibinfo{editor}{X.~Leoncini}, and \bibinfo{editor}{G.~Zaslavsky}, eds.,
  \emph{\bibinfo{booktitle}{Chaos, Complexity, and Transport: Theory and
  Applications}}, \bibinfo{pages}{55--68} (\bibinfo{publisher}{World
  Scientific}, \bibinfo{address}{Singapore}, \bibinfo{year}{2008}).

\bibitem[{Boffetta \emph{et~al.}(2009)Boffetta, {De Lillo}, and
  Mazzino}]{Boffetta2009}
\bibinfo{author}{G.~Boffetta}, \bibinfo{author}{F.~{De Lillo}}, and
  \bibinfo{author}{A.~Mazzino}, \enquote{\bibinfo{title}{Peripheral mixing of
  passive scalar at small {R}eynolds number},} \emph{\bibinfo{journal}{J. Fluid
  Mech.}} \textbf{\bibinfo{volume}{624}}, \bibinfo{pages}{151--158}
  (\bibinfo{year}{2009}), \eprint{arXiv:0811.4519}.

\bibitem[{Zaggout and Gilbert(2011)}]{Zaggout2011}
\bibinfo{author}{F.~A. Zaggout} and \bibinfo{author}{A.~D. Gilbert},
  \enquote{\bibinfo{title}{Passive scalar decay in chaotic flows with
  boundaries},}  (\bibinfo{year}{2011}), \bibinfo{note}{preprint}.

\bibitem[{Gouillart \emph{et~al.}(2009)Gouillart, Dauchot, Thiffeault, and
  Roux}]{Gouillart2009}
\bibinfo{author}{E.~Gouillart}, \bibinfo{author}{O.~Dauchot},
  \bibinfo{author}{J.-L. Thiffeault}, and \bibinfo{author}{S.~Roux},
  \enquote{\bibinfo{title}{Open-flow mixing: Experimental evidence for strange
  eigenmodes},} \emph{\bibinfo{journal}{Phys. Fluids}}
  \textbf{\bibinfo{volume}{21}}~(\bibinfo{number}{2}), \bibinfo{pages}{022603}
  (\bibinfo{year}{2009}).

\bibitem[{Gouillart \emph{et~al.}(2010)Gouillart, Thiffeault, and
  Dauchot}]{Gouillart2010}
\bibinfo{author}{E.~Gouillart}, \bibinfo{author}{J.-L. Thiffeault}, and
  \bibinfo{author}{O.~Dauchot}, \enquote{\bibinfo{title}{Rotation shields
  chaotic mixing regions from no-slip walls},} \emph{\bibinfo{journal}{Phys.
  Rev. Lett.}} \textbf{\bibinfo{volume}{104}}~(\bibinfo{number}{20}),
  \bibinfo{pages}{204502} (\bibinfo{year}{2010}).

\bibitem[{Finn \emph{et~al.}(2003)Finn, Cox, and Byrne}]{MattFinn2003}
\bibinfo{author}{M.~D. Finn}, \bibinfo{author}{S.~M. Cox}, and
  \bibinfo{author}{H.~M. Byrne}, \enquote{\bibinfo{title}{Topological chaos in
  inviscid and viscous mixers},} \emph{\bibinfo{journal}{J. Fluid Mech.}}
  \textbf{\bibinfo{volume}{493}}, \bibinfo{pages}{345--361}
  (\bibinfo{year}{2003}).

\bibitem[{Reichl(2004)}]{ReichlTransition}
\bibinfo{author}{L.~E. Reichl}, \emph{\bibinfo{title}{The Transition to Chaos:
  Conservative Classical Systems and Quantum Manifestations}},
  \bibinfo{edition}{second} edition (\bibinfo{publisher}{Springer},
  \bibinfo{address}{New York}, \bibinfo{year}{2004}).

\bibitem[{{El Omari} and {Le Guer}(2009)}]{ElOmari2009}
\bibinfo{author}{K.~{El Omari}} and \bibinfo{author}{Y.~{Le Guer}},
  \enquote{\bibinfo{title}{Numerical study of thermal chaotic mixing in a two
  rod rotating mixer},} \emph{\bibinfo{journal}{Comput. Therm. Sci.}}
  \textbf{\bibinfo{volume}{1}}~(\bibinfo{number}{1}), \bibinfo{pages}{55--73}
  (\bibinfo{year}{2009}).

\bibitem[{{El Omari} and {Le Guer}(2010)}]{ElOmari2010}
\bibinfo{author}{K.~{El Omari}} and \bibinfo{author}{Y.~{Le Guer}},
  \enquote{\bibinfo{title}{Alternate rotating walls for thermal chaotic
  mixing},} \emph{\bibinfo{journal}{Int. J. Heat Mass Transfer}}
  \textbf{\bibinfo{volume}{53}}~(\bibinfo{number}{1-3}),
  \bibinfo{pages}{123--134} (\bibinfo{year}{2010}).

\bibitem[{Thiffeault \emph{et~al.}(2008)Thiffeault, Finn, Gouillart, and
  Hall}]{Thiffeault2008b}
\bibinfo{author}{J.-L. Thiffeault}, \bibinfo{author}{M.~D. Finn},
  \bibinfo{author}{E.~Gouillart}, and \bibinfo{author}{T.~Hall},
  \enquote{\bibinfo{title}{Topology of chaotic mixing patterns},}
  \emph{\bibinfo{journal}{Chaos}} \textbf{\bibinfo{volume}{18}},
  \bibinfo{pages}{033123} (\bibinfo{year}{2008}), \eprint{arXiv:0804.2520}.

\bibitem[{Finn and Cox(2001)}]{MattFinn2001}
\bibinfo{author}{M.~D. Finn} and \bibinfo{author}{S.~M. Cox},
  \enquote{\bibinfo{title}{{S}tokes flow in a mixer with changing geometry},}
  \emph{\bibinfo{journal}{J. Eng. Math.}}
  \textbf{\bibinfo{volume}{41}}~(\bibinfo{number}{1}), \bibinfo{pages}{75--99}
  (\bibinfo{year}{2001}).

\bibitem[{Clifford \emph{et~al.}(2004)Clifford, Cox, and Finn}]{Clifford2004}
\bibinfo{author}{M.~J. Clifford}, \bibinfo{author}{S.~M. Cox}, and
  \bibinfo{author}{M.~D. Finn}, \enquote{\bibinfo{title}{{R}eynolds number
  effects in a simple planetary mixer},} \emph{\bibinfo{journal}{Chem. Eng.
  Sci.}} \textbf{\bibinfo{volume}{59}}~(\bibinfo{number}{16}),
  \bibinfo{pages}{3371--3379} (\bibinfo{year}{2004}).

\bibitem[{Stroock \emph{et~al.}(2002)Stroock, Dertinger, Ajdari, Mezi\'{c},
  Stone, and Whitesides}]{Stroock2002}
\bibinfo{author}{A.~D. Stroock}, \bibinfo{author}{S.~K.~W. Dertinger},
  \bibinfo{author}{A.~Ajdari}, \bibinfo{author}{I.~Mezi\'{c}},
  \bibinfo{author}{H.~A. Stone}, and \bibinfo{author}{G.~M. Whitesides},
  \enquote{\bibinfo{title}{Chaotic mixer for microchannels},}
  \emph{\bibinfo{journal}{Science}} \textbf{\bibinfo{volume}{295}},
  \bibinfo{pages}{647--651} (\bibinfo{year}{2002}).

\bibitem[{Villermaux \emph{et~al.}(2008)Villermaux, Stroock, and
  Stone}]{Villermaux2008}
\bibinfo{author}{E.~Villermaux}, \bibinfo{author}{A.~D. Stroock}, and
  \bibinfo{author}{H.~A. Stone}, \enquote{\bibinfo{title}{Bridging kinematics
  and concentration content in a chaotic micromixer},}
  \emph{\bibinfo{journal}{Phys. Rev. E}}
  \textbf{\bibinfo{volume}{77}}~(\bibinfo{number}{1}), \bibinfo{pages}{15301}
  (\bibinfo{year}{2008}).

\bibitem[{Abdullaev(2006)}]{Abdullaev}
\bibinfo{author}{S.~S. Abdullaev}, \emph{\bibinfo{title}{Construction of
  Mappings for Hamiltonian Systems and Their Applications}},
  \emph{\bibinfo{series}{Lecture Notes in Physics}}, volume
  \bibinfo{volume}{691} (\bibinfo{publisher}{Springer},
  \bibinfo{address}{Berlin}, \bibinfo{year}{2006}).

\end{thebibliography}

\end{document}